\documentclass[12pt,a4paper]{article}
\usepackage[utf8]{inputenc}
\usepackage{a4wide}
\usepackage{hyperref}
\usepackage{graphicx}
\newcommand{\be}{\begin{equation}}
\newcommand{\ee}{\end{equation}}
\newcommand{\ba}{\begin{eqnarray}}
\newcommand{\ea}{\end{eqnarray}}

\newcommand{\lgl}{\langle}
\newcommand{\rgl}{\rangle}
\newcommand{\trc}[1]{\mathrm{tr}_c\left(#1\right)}
\newcommand{\trf}[1]{\mathrm{tr}_F\left(#1\right)}
\newcommand{\one}{\mathrm{I}}
\begin{document}

\begin{titlepage}
\begin{flushright}
LU TP 15-34\\
September 2015
\end{flushright}
\vfill
\begin{center}
{\Large\bf Finite Volume and Partially Quenched\\[4mm]QCD-like Effective Field Theories}

\vfill

{\bf Johan Bijnens and Thomas Rössler}\\[0.3cm]
{Department of Astronomy and Theoretical Physics, Lund University,\\
S\"olvegatan 14A, SE 223-62 Lund, Sweden}
\end{center}
\vfill
\begin{abstract}
We present a calculation of the meson masses,
decay constants  and quark-antiquark vacuum expectation value
for the three generic QCD-like chiral symmetry breaking patterns
$SU(N_F)\times SU(N_F)\to SU(N_F)_V$, $SU(N_F)\to SO(N_F)$ and
$SU(2N_F)\to Sp(2 N_F)$ in the effective field theory for these cases.
We extend the previous two-loop work to include effects of partial quenching
and finite volume.

The calculation has been
performed using the  quark flow technique.
We reproduce the known infinite volume results in the unquenched case.
The analytical results can be found in the supplementary material.

Some examples of numerical results are given.
The numerical programs for all cases are included
in version 0.54 of the \textsc{CHIRON} package.

The purpose of this work is the use in lattice extrapolations to
zero mass for QCD-like and strongly interacting Higgs sector lattice
calculations.
\end{abstract}
\vfill
\vfill
\end{titlepage}


\section{Introduction}
\label{sec:Introduction}

Effective field theory is used extensively in the study of strongly
interacting gauge theories. A recent review covering a number of different
applications in addition to other methods is \cite{Brambilla:2014jmp}.
Besides general interest in understanding strongly interacting gauge theories,
they might still be useful as an alternative for the Standard Model
Higgs sector as well as for dark matter.
These applications have been reviewed recently at the 2015
\cite{talk1lattice2015,talk2lattice2015} and 2013 \cite{Kuti:2014epa}
lattice conferences.
A number of recent lattice studies is \cite{lattice}. Reviews of
technicolor and strongly interacting Higgs sectors are
\cite{Andersen:2011yj,Techni1,Techni2}.

Lattice studies are always performed at a nonzero fermion mass. In order
to obtain results in the massless limit extrapolations are needed.
A main tool for this in the context of lattice QCD is Chiral
Perturbation Theory (ChPT) \cite{Weinberg:1978kz,Gasser:1983yg,Gasser:1984gg}.

In the case of equal mass fermions three main symmetry breaking patterns
are possible \cite{Peskin:1980gc,Preskill:1980mz,Dimopoulos:1979sp}.
For $N_F$ Dirac fermions in a complex representation the
global symmetry group is $SU(N_F)_L\times SU(N_F)_R$
and it breaks spontaneously to the diagonal subgroup $SU(N_F)_V$.
For $N_F$ Dirac fermions in a real representation the
global symmetry group is $SU(2N_F)$
and it breaks spontaneously to $SO(2N_F)$.
An alternative possibility is that we have $N_F$ Majorana fermions
in a real representation with a global symmetry group $SU(N_F)$ spontaneously
broken to $SO(N_F)$. We show in this work that the EFT for the quantities
we consider is really the same as for Dirac fermions.
The final case is $N_F$ Dirac fermions in a pseudo-real representation.
The global symmetry group is again $SU(2N_F)$ but in this case it is expected
to be broken spontaneously to $Sp(2N_F)$.

The effective field
theory (EFT) for these cases is discussed at tree level or lowest order (LO)
in \cite{Kogut}. At next-to-leading order (NLO) the
first case is simply ChPT for $N_F$ light quarks with a symmetry breaking
pattern of $SU(N_F)\times SU(N_F)\to SU(N_F)$, a direct extension of the
QCD case and was already done in \cite{Gasser:1984gg}. The pseudo-real
case was done
at NLO by \cite{Splittorff}. The $SU(2N_F)\to SO(2N_F)$ case was done
in \cite{Toublan:1999hi}. The extension for all three cases to
next-to-next-to-leading order (NNLO) was done
in earlier work by one of the authors \cite{Bijnens:2009qm}. More references
to earlier work can be found there and in \cite{Bijnens:2011fm,Bijnens:2011xt}.

This paper is an extension to the work of \cite{Bijnens:2009qm}. We add a short
discussion showing that the calculations and the Lagrangian for the real
case also covers the case of Majorana fermions in a real representation.
The main part of the work concerns the extension of the calculations
at NNLO order of the masses, decay constants and vacuum expectation values
to include effects of partial quenching and finite volume.

Partial quenching was introduced in ChPT by \cite{Bernard:1993sv}.
A thorough discussion of the assumptions involved is in
\cite{Bernard:2013kwa}. It allows to study a number of variations
of input parameters at reduced cost, as discussed in e.g.
\cite{Sharpe:2000bc}. We do not use the supersymmetric method
introduced in \cite{Bernard:1993sv} and extended (at NLO) to the
cases discussed here in \cite{Toublan:1999hi}. 
We only use the quark-flow technique introduced in \cite{Sharpe:1992ft}.
Two-loop results in infinite volume  partially quenched ChPT (PQChPT)
for the masses and decay constants are in
\cite{Bijnens:2004hk,Bijnens:2005ae,Bijnens:2006jv}. The definitions of
the infinite volume integrals we use can be found there.

Finite volume effects in ChPT were introduced in ChPT in
\cite{Gasser:1986vb,Gasser:1987ah,Gasser:1987zq}.
Early two-loop work is
\cite{Colangelo:2006mp,Bijnens:2006ve}. The vacuum expectation
value was discussed in more detail in \cite{Damgaard:2008zs}.
After the proper evaluation of the finite volume two-loop
sunsetintegrals using two different methods \cite{Bijnens:2013doa}
the masses and decay constants were treated
in both the unquenched \cite{Bijnens:2014dea}
and partially quenched \cite{Bijnens:2015dra} case.
In particular the integral notation at finite volume we use is defined 
in \cite{Bijnens:2015dra}.

In Sect.~\ref{quarklevel} we recapitulate briefly the discussion from
\cite{Bijnens:2009qm} at the quark level and add the case with Majorana
fermions. Sect.~\ref{EFT} similarly recapitulates \cite{Bijnens:2009qm}
at the effective field theory level and adds the Majorana fermion case.
The cases with Dirac fermions and Majorana fermions are essentially
identical from the EFT point of view for the quantities we consider.
The underlying reason is an $U(2N_F)$ transformation that relates the
two cases as discussed in Sect.~\ref{majoranadirac}.
Partial quenching and the quark.flow techniques we have used for the
different cases is discussed to some extent in Sect.~\ref{PQCHPT}.
For a discussion on finite volume and the notation used there we
refer to \cite{Bijnens:2015dra}.
Our analytical results are described in Sect.~\ref{analytical}, in particular
we clarify the definitions of the decay constant and vacuum expectation value
used in terms of quark fields.
The numerical examples and checks are presented in
Sect.~\ref{numerical}. The analytical formulas are included in the
supplementary file \cite{analyticalresults} and the numerical programs
are available via \textsc{CHIRON},
\cite{Bijnens:2014gsa,chiron}.
The last section briefly recapitulates the main points of our work.

\section{Quark level}
\label{quarklevel}

\subsection{The three Dirac fermion cases}

The discussion here is kept very short, longer versions can be found
in \cite{Kogut} and \cite{Bijnens:2009qm}. This subsection is mainly
included to show normalization conventions.

\paragraph{QCD or complex representation}

In the $N_F$ equal mass Dirac fermions in a complex representation, we put the
$N_F$ fermions together in an $N_F$ column matrix $q$.
The global symmetry transformation by $g_L\times g_R \in SU(N_F)_L\times
SU(N_F)_R$ is given by
\be
\label{transformQCD}
q_L\to g_L q_L,~~ q_R\to g_R q_R,
l_\mu \to g_L l_\mu g_L^\dagger +i g_L\partial_\mu g_L^\dagger\,,~~
r_\mu \to g_R l_\mu g_R^\dagger +i g_R\partial_\mu g_R^\dagger\,,~~
\mathcal{M}\to g_R\mathcal{M}g_L^\dagger\,.
\ee
The matrix $\mathcal{M}=m_q\one+s+ip$ brings the quark mass term $m_q \one$
and the external scalar $s$ and pseudo-scalar densities
in the Lagrangian via $-\overline q_R\mathcal{M}q_L+\mathrm{h.c.}$.
The external fields $l_\mu,r_\mu$ are in the Lagrangian via
$\overline q_L \gamma^\mu l_\mu q_L+\overline q_R \gamma^\mu r_\mu q_R$.
Taking derivatives w.r.t. the external fields allows to calculate relevant
Green functions \cite{Gasser:1983yg,Gasser:1984gg}. In particular, deriving w.r.t. $s_{11}$
allows us to obtain $\lgl\overline q_{L1}q_{R1}+\overline q_{R1}q_{L1}\rgl$
and derivatives w.r.t. $a_{\mu12}$ with $r_{\mu12}=-l_{\mu12}=a_{\mu12}$
allows access to matrix-elements of $\overline q_2 \gamma^\mu\gamma^5 q_1$ 
The symmetry is spontaneously broken by a vacuum expectation value
\be
\lgl\overline q_{Lj} q_{Ri}\rgl = v_0 \delta_{ij}\,.
\ee
This leaves a global symmetry $SU(N_F)_V$ with $g_L=g_R$ unbroken.

\paragraph{Adjoint or real representation}

When the fermions are in a real representation, we can introduce besides
the $N_F$ right handed fermions $q_{Ri}$ a second set of right handed fermions
in the same gauge group representation, $\tilde q_{Ri} = C\overline q_{Li}^T$.
These can be put together in a $2N_F$ column vector $\hat q$,
$\hat q^T = (q_{R1}\ldots q_{RN_F}~\tilde q_{R1}\ldots\tilde q_{RN_F})$.
The global symmetry transformation with $g\in SU(2N_F)$ is now
\be
\hat q\to g \hat q,~~\hat V_\mu\to g\hat V_\mu g^\dagger,~~
\hat\mathcal{M}\to g\hat\mathcal{M}g^T.
\ee
We define the external densities and currents as in the QCD case with
$r_\mu,l_\mu$ and $\mathcal M$. We define $2N_F\times2N_F$ matrices
\be
\hat\mathcal{M} = 
\left(\begin{array}{cc} 0 & \mathcal{M}\\\mathcal{M}^T & 0 \end{array}\right),
~~
\hat V_\mu = 
\left(\begin{array}{cc} r_\mu& 0\\ 0 & -l_\mu^T \end{array}\right)\,.
\ee
Note that the global symmetry can change quark-antiquark currents to
diquark currents. The fermions condense forming a vacuum expectation value
\be
\frac{1}{2}\lgl\hat q^T_j C \hat q_j\rgl
= v_0 J_{Sij}\,
\quad
J_S =\left(\begin{array}{cc} 0 & \one\\ \one & 0\end{array}\right)\,.
\ee
This leaves a global symmetry $SO(2N_F)$ with $g J_S g^T = \one$.

\paragraph{$N_c=2$ or pseudo-real representation}

When the fermions are in a pseudo-real representation,
we can introduce besides
the $N_F$ right handed fermions $q_{Ria}$
again a second set of right handed fermions
in the same gauge group representation,
$\tilde q_{Ria} =\epsilon_{ab} C\overline q_{Lib}^T$.
$a,b$ are gauge indices and the extra Levi-Civita tensor $\epsilon_{ab}$
is needed to have $\tilde q_{Ria}$ transform under the gauge group
as $q_{iRa}$. The explicit formula is for the case of the fundamnetal
representation with $N_c=2$.
$q_{Ri}$ and $\tilde q_{Ri}$ can be put together in a $2N_F$ column vector
$\hat q$,
$\hat q^T = (q_{R1}\ldots q_{RN_F}~\tilde q_{R1}\ldots\tilde q_{RN_F})$.
The global symmetry transformation with $g\in SU(2N_F)$ is now
\be
\hat q\to g \hat q,~~\hat V_\mu\to g\hat V_\mu g^\dagger,~~
\hat\mathcal{M}\to g\hat\mathcal{M}g^T.
\ee
We define the external densities and currents as in the QCD case with
$r_\mu,l_\mu$ and $\mathcal M$. We then define
\be
\hat\mathcal{M} = 
\left(\begin{array}{cc} 0 & -\mathcal{M}\\\mathcal{M}^T & 0 \end{array}\right),
~~
\hat V_\mu = 
\left(\begin{array}{cc} r_\mu& 0\\ 0 & -l_\mu^T \end{array}\right)\,.
\ee
Note that the global symmetry can again change quark-antiquark currents to
diquark currents. The fermions condense forming a vacuum expectation value
\be
\frac{1}{2}\lgl\hat q^T_{ja}\epsilon_{ab} C \hat q_{jb}\rgl
= v_0 J_{Aij}\,
\quad
J_A =\left(\begin{array}{cc} 0 & -\one\\ \one & 0\end{array}\right)\,.
\ee
This leaves a global symmetry $Sp(2N_F)$ with $g J_A g^T = \one$.

\subsection{Majorana fermions in a real representation}

In the earlier work \cite{Bijnens:2009qm}
at infinite volume Dirac fermions and Dirac masses were
assumed. It was then also asumed that the vacuum condensate was aligned with
the Dirac fermion masses.
There is in fact another possibility. Majorana fermions with a Majorana mass
in a real representation of the gauge group. In this case the global symmetry
is $SU(N_F)$. It is expected to be spontaneously broken down to $SO(N_F)$
which is aligned with the Majorana masses.

A Majorana spinor is a Dirac spinor that satisfies
\be
\label{defmajorana}
\psi = C\overline\psi^T\,\quad
\mathrm{or}\quad
\psi =
\left(\begin{array}{c} \psi_M \\ -i\sigma^2 \psi_M^*\end{array}\right)\,.
\ee
The last equality are in the chiral representation for the Dirac matrices.
The Lagrangian for a single free Majorana fermion is
\be
\label{lagMajorana}
\frac{1}{2}
\overline\psi i\gamma^\mu\partial_\mu\psi-\frac{m}{2}\overline\psi\psi
 = \psi_M^\dagger C i\overline\sigma^\mu\partial_\mu \psi
-\frac{im}{2}\left(\psi^T_M\sigma^2\psi+\psi_M^\dagger\sigma^2\psi^*\right)\,.
\ee
$\overline\sigma^0=\one,\overline\sigma^i = -\sigma^i$.
If we want to gauge this for $m\ne0$ the mass term requires the fermions to be
in a real representation of the gauge group.

For $N_F$ Majorana fermions $\psi_{Mi}$ in the adjoint representation with
external fields $\hat V_\mu$ and $\hat\mathcal{M}$ the Lagrangian,
put in a big column vector $\hat q^T =( \psi_1^T\ldots\psi_{N_F}^T)$
is
\be
\mathcal{L} = \frac{1}{2}\trc{\hat q^\dagger i\overline\sigma^\mu(i D_\mu+\hat V_\mu) \hat q}
-\frac{1}{2}\trc{\hat q^T \sigma^2 \hat\mathcal{M}^\dagger \hat q+\hat q^\dagger \sigma^2 \hat\mathcal{M} \hat q^*}\,.
\ee
This Lagrangian has a global $SU(N_F)$ symmetry with $g\in SU(N_F)$ with
\be
\hat q\to g \hat q,,\quad 
\hat V_\mu\to g \hat V_\mu g^\dagger+i g\partial_\mu g^\dagger\,,
\quad
\hat\mathcal{M}\to g\hat\mathcal{M}g^T\,.
\ee

The maximal symmetry argument says that in this case the fermions will
condense to the flavour neutral vacuum
$\lgl\trc{\hat q^T C \hat q}\rgl$.
This is conserved by the part of the global group that satisfies $gg^T=\one$
or the conserved part of the global symmetry group is $SO(N_F)$.

Note that the form of the vacuum and the form of the mass term
are the only differences as far as the global
symmetry group and its breaking
are concerned compared to the case with $N_F/2$ Dirac fermions
in a real representation.

\section{Effective field theory}
\label{EFT}

\subsection{The general LO and NLO Lagrangian}

The ChPT Lagrangian for $N_F$ flavours at LO and NLO has been derived in
\cite{Gasser:1984gg}. The Lagrangian for the other cases has the same form as
has been shown in \cite{Kogut,Bijnens:2009qm} and other papers.
The precise derivation can be found in \cite{Bijnens:2009qm}
and the Majorana fermion case below in Sect.~\ref{EFTmajorana}.

In terms of the quantities $u_\mu,f_{\pm\mu\nu},\chi_{\pm}$ defined below for
each case the lowest order Lagangian is
\be
\label{L2}
\mathcal{L}_2 = \frac{F^2}{4}\langle u_\mu u^\mu +\chi_+\rangle\,.
\ee
Here we use the notation $\lgl A \rgl = \trf{A}$, denoting the trace over
flavours.
The NLO Lagrangian derived by \cite{Gasser:1984gg} reads
\ba
\label{L4}
\mathcal{L}_4 &=&
L_0 \langle u^\mu u^\nu u_\mu u_\nu \rangle
+L_1 \langle u^\mu u_\mu\rangle\langle u^\nu u_\nu \rangle
+L_2 \langle u^\mu u^\nu\rangle\langle u_\mu u_\nu \rangle
+L_3 \langle u^\mu u_\mu u^\nu u_\nu \rangle
\nonumber\\&&
+L_4  \langle u^\mu u_\mu\rangle\langle\chi_+\rangle
+L_5  \langle u^\mu u_\mu\chi_+\rangle
+L_6 \langle\chi_+\rangle^2
+L_7 \langle\chi_-\rangle^2
+\frac{1}{2} L_8 \langle\chi_+^2+\chi_-^2\rangle
\nonumber\\&&
-i L_9\langle f_{+\mu\nu}u^\mu u^\nu\rangle
+\frac{1}{4}L_{10}\langle f_+^2-f_-^2\rangle
+\frac{1}{2}H_1\langle f_+^2-f_-^2\rangle
+\frac{1}{4}H_2\langle\chi_+^2-\chi_-^2\rangle\,.
\ea
The NNLO Lagrangian has been classified for the $N_F$-flavour case
in \cite{Bijnens:1999sh}. The Lagrangian at NNLO for the other cases is
not known, the direct equivalent of the results in  \cite{Bijnens:1999sh}
is definitely a complete Lagrangian but might not be minimal.
For this reason we do not quote the dependence on the NNLO Lagrangian in the
real and pseudo-real cases.

The divergences at NLO were derived for the QCD case in
\cite{Gasser:1984gg}, for the others in \cite{Splittorff,Bijnens:2009qm}.
At NNLO only the QCD case is known \cite{Bijnens:1999hw}.

\subsection{The three Dirac fermion cases}

A more extensive discussion can be found in \cite{Kogut,Bijnens:2009qm}.
Here we simply quote the results.

When we have a global symmetry group $G$ with generators $T^a$ which
is spontaneously broken down to a subgroup $H$ with generators $Q^a$
which form a subset of the $T^a$, the Goldstone bosons can be described by the
coset $G/H$. This coset can be parametrized \cite{CCWZ}
via the broken generators $X^a$. Below we explain what is used for the
different cases. 
We always work with generators normalized to 1, i.e.
$\lgl X^a X^b\rgl=\delta^{ab}$.

The quantities used from the quark level are given in Sect.~\ref{quarklevel}.

\paragraph{QCD or complex representation}
The Goldstone boson manifold is in this case
$SU(N_F)\times SU(N_F)/SU(N_F)$ which itself has the structure of an $SU(N_F)$
Note that the axial generators do not generate a subgroup of
$SU(N_F)\times SU(N_F)$ even if $G/H$ has the structure of a group in this
case.

We choose as the broken generators $X^a$ the generators of $SU(N_F)\approx G/H$.
The quantities needed to construct the Lagrangian and their symmetry
transformations are
\ba
u &=& \exp\left(\frac{i}{\sqrt{2}F}\pi^a X^a\right)
\to g_R u h^\dagger\equiv h u g_L^\dagger
\nonumber\\
u_\mu &=&i\left(u^\dagger(\partial_\mu-ir_\mu)u
-u(\partial_\mu-l_\mu)u^\dagger\right)\to h u_\mu h^\dagger\,,
\nonumber\\
\chi &=& 2 B_0 \mathcal{M} \to g_R \chi g_L^\dagger 
\nonumber\\
\chi_\pm &=& u^\dagger \chi u^\dagger\pm u\chi^\dagger u
\to h \chi_\pm h^\dagger\,,
\nonumber\\
l_{\mu\nu}&=& \partial_\mu l_\nu-\partial_\mu l_\nu -il_\mu l_\mu+il_\nu l_\mu
  \to g_L l_{\mu\nu} g_L^\dagger
\nonumber\\
r_{\mu\nu}&=& \partial_\mu r_\nu-\partial_\mu r_\nu -ir_\mu r_\mu+ir_\nu r_\mu
  \to g_R r_{\mu\nu} g_R^\dagger
\nonumber\\
f_{\pm\mu\nu}&=& u l_{\mu\nu} u^\dagger \pm u^\dagger r_{\mu\nu} u
\to h f_{\pm\mu\nu} h^\dagger\,.
\ea
The first line defines $h$ \cite{CCWZ}. 

\paragraph{Adjoint or real representation}
The Goldstone boson manifold is in this case $SU(2N_F)/SO(2N_F)$.
The unbroken generators satisfy $ Q^a J_S = -J_S Q^{aT}$ which follows from
$g J_S g^T = J_S$. The broken generators satisfy
$J_S X^a = X^{aT} J_S$.

The quantities needed to construct the Lagrangians are \cite{Bijnens:2009qm}
\ba
u &=& \exp\left(\frac{i}{\sqrt{2}F}\pi^a X^a\right)
\to g u h^\dagger
\nonumber\\
u_\mu &=& i\left(u^\dagger(\partial_\mu-i\hat V_\mu)u
 - u(\partial_\mu+iJ_S\hat V^T_\mu J_S)u^\dagger\right)\,,
\nonumber\\
\chi&=& 2 B_0 \hat\mathcal{M}
\nonumber\\
\chi_\pm &=& u^\dagger \chi J_S u^\dagger\pm u J_S \chi^\dagger u
\nonumber\\
\hat V_{\mu\nu}&=&\partial_\mu \hat V_\nu-\partial_\nu\hat V_\mu-i
\left(\hat V_\mu\hat V_\nu-\hat V_\nu\hat V_\mu\right)
\nonumber\\  
f_{\pm\mu\nu} &=& J_S u\hat V_{\mu\nu} u^\dagger J_S\pm  u\hat V_{\mu\nu} u^\dagger
\ea
The first line defines $h$ by requiring that $guh^\dagger$ is
of the form $\exp(i\pi^aX^a/(\sqrt{2}F))$.
Note that the derivation used $J_S u = u^T J_S$.

\paragraph{$N_c=2$ or pseudo-real representation}
The Goldstone boson manifold is $SU(2N_F)/Sp(2N_F)$.
The unbroken generators satisfy $ Q^a J_A = -J_A Q^{aT}$ which follows from
$g J_A g^T = J_A$. The broken generators satisfy
$J_A X^a = X^{aT} J_A$.

The quantities needed are \cite{Bijnens:2009qm}
\ba
u &=& \exp\left(\frac{i}{\sqrt{2}F}\pi^a X^a\right)
\to g u h^\dagger
\nonumber\\
u_\mu &=& i\left(u^\dagger(\partial_\mu-i\hat V_\mu)u
 - u(\partial_\mu+iJ_A\hat V^T_\mu J_A^T)u^\dagger\right)\,,
\nonumber\\
\chi&=& 2 B_0 \hat\mathcal{M}
\nonumber\\
\chi_\pm &=& u^\dagger \chi J_A^T u^\dagger\pm u J_A \chi^\dagger u
\nonumber\\
\hat V_{\mu\nu}&=&\partial_\mu \hat V_\nu-\partial_\nu\hat V_\mu-i
\left(\hat V_\mu\hat V_\nu-\hat V_\nu\hat V_\mu\right)
\nonumber\\  
f_{\pm\mu\nu} &=& J_A u\hat V_{\mu\nu} u^\dagger J_A^T\pm  u\hat V_{\mu\nu} u^\dagger
\ea
The first line defines $h$ by requiring that $guh^\dagger$ is
of the form $\exp(i\pi^aX^a/(\sqrt{2}F))$.
Note that the derivation used $J_A u = u^T J_A$.

\subsection{Majorana fermions in a real representation}
\label{EFTmajorana}

The vacuum in this case is characterized by the condensate
\be
\frac{1}{2}\langle \hat q_{i}^T C \hat q_j\rangle
= \frac{1}{2}\langle\overline q q\rangle\delta_{ij}
\,.
\ee
Under the symmetry group $g\in SU(N_F)$ this moves around as
\be
\delta{ij} \to \left(g^T g\right)_{ij}\,.
\ee
The unbroken part of the group is given by the generators $\tilde Q^a$
and the broken part by the generators $\tilde X^a$ which satisfy
\be
\label{commutatorsAdjoint2}
\tilde Q^a = -\tilde Q^{aT}\,,\qquad\tilde X^a =\tilde X^{aT}\,.
\ee
Just as in the cases discussed in \cite{Bijnens:2009qm}
we can construct a rotated vacuum
in general by using the broken part of the symmetry group
on the vacuum. This leads to a matrix
\be
U = u u^T\to g U g^T\qquad\mathrm{with}\qquad
u=\exp\left(\frac{i}{\sqrt{2}\,F}\pi^a X^a\right)\,.
\ee
The matrix $u$ transforms as in the general $CCWZ$ case
as
\be
u \to g u h^\dagger\,.
\ee
Some earlier work used the matrix $U$ to describe
the Lagrangian \cite{Kogut}. Here we will, as in \cite{Bijnens:2009qm}
use the CCWZ scheme
to obtain a notation that is formally identical to the
QCD case. We add $N_F\times N_F$ matrices of external fields $\hat V_\mu$
and $\hat\mathcal{M}$. We need to obtain the $u_\mu$, or broken generator,
parts of $u^\dagger\left(\partial_\mu-iV_\mu\right)u$.
Eq.~(\ref{commutatorsAdjoint2}) have as a consequence that
$u$ satisfies
\be
\label{JSu2}
u  = u^T \,.
\ee
This leads using the same method as in \cite{Bijnens:2009qm} to
\ba
\label{defumuadjoint2}
u_\mu &=& i\left(u^\dagger(\partial_\mu-i\hat V_\mu)u
 - u(\partial_\mu+i\hat V^T_\mu )u^\dagger\right)\,.
\ea
With this we can construct Lagrangians.
The equivalent quantities to
the field strengths are
\be
\label{deffieldadjoint2}
f_{\pm\mu\nu} =  u\hat V_{\mu\nu} u^\dagger \pm  u\hat V_{\mu\nu} u^\dagger
\ee
with $\hat V_{\mu\nu}=\partial_\mu \hat V_\nu-\partial_\nu\hat V_\mu-i
\left(\hat V_\mu\hat V_\nu-\hat V_\nu\hat V_\mu\right)$ and for the mass matrix
\be
\label{defchiadjoint2}
\chi_\pm = u^\dagger\chi u^{\dagger T} \pm  u^T\chi^\dagger u
\ee
with $\chi = 2B_0 \hat\mathcal{M}$.
The Lagrangians at LO and NLO have exactly the same
form as given in (\ref{L2}) and (\ref{L4}) with
$u_\mu$, $\chi_\pm$ and $f_{\pm\mu\nu}$
as defined in (\ref{defumuadjoint2}), (\ref{deffieldadjoint2})
and (\ref{defchiadjoint2}).

\section{Relation Dirac and Majorana for the adjoint case}
\label{majoranadirac}

As discussed below, we have calculated the adjoint case using two methods.
They were appropriate for the Dirac and the Majorana case respectively.
After doing the trivial $2N_F\to N_F$ change the results agreed exactly.
If we compare the two cases, we see that the main difference is really the
choice of vacuum.

The Dirac and Majorana cases lead to a choice of vacuum
\be
\lgl\hat q^T_i C \hat q_j\rgl_D \propto J_{Sij}
\,,\quad
\lgl\hat q^T_i C \hat q_j\rgl_D \propto \one_{ij}\,.
\ee
Is it possible to relate the two cases in a simple way?
Under a global symmetry transformation the first one transforms as
$J_S\to g J_S g^T$. If we could find a global transformation $g_R$
that lead to $g_R J_S g_R^T=\one$ the two cases would be obviously the same.

It is not possible in general with a $SU(2N_F)$ rotation to accomplish this
since $\det J_S=\pm1$ ($-1$ for the $2N_F=2$) while $\det\one=1$.
However it is possible with a $U(2N)$ transformation.
An explicit choice for $g_R$, with a free phase $\alpha$ is
\be
g_R = \frac{1}{\sqrt{2}}\left(\begin{array}{cc}
\mp i e^{i\alpha}\one & \pm i e^{-i\alpha} \one\\
e^{i\alpha} \one & e^{-i\alpha}\one\end{array}\right)\,. 
\ee
It can be checked that this transforms a Dirac mass term for $N_F$ Dirac
fermions into a Majorana mass term for $2N_F$ Majorana fermions.

Inspections of the effective Lagrangians needed lead to the immediate
conclusion that the mass independent terms really are $U(2N_F)$ invariant,
and the mass dependent terms for the two cases are turned into each other.

$g_R$ can also be used to relate the two different embeddings of $SO(2N_F)$
in $SU(2N_F)$ to each other. For the Dirac case the $SO(2N)$ generators
satisfied $Q^{aT} J_S = -J_S Q^a$
while for the Majorana case they satisfied
$\tilde Q^{aT}=-\tilde Q^a$. The two sets of generators are related by
\be
\tilde Q^a = g_R Q^a g_R^\dagger\,,\quad
\tilde X^a = g_R X^a g_R^\dagger\,.
\ee

\section{Partially quenching and the quark flow technique}
\label{PQCHPT}

A thorough discussion of PQChPT and in particular the derivation of
the propagator used there is \cite{Sharpe:2001fh}.
That discussion uses the supersymmetric method. Alternative methods of
calculation are the replica trick \cite{Damgaard:2000gh} and the quark flow
method \cite{Sharpe:1992ft}. The earliest partially quenched work for
QCDlike theories used the supersymmetric method \cite{Toublan:1999hi}.
The replica trick has been used in \cite{Levinsen:2003be}.
We use the quark-flow method.

For this method we look at the matrix
\be
\Phi = \pi^a X^a
\ee
for each of the cases. 

For the QCD case, $\Phi$ is a traceless Hermitian matrix.
We actually keep $\Phi$
in the flavour basis with elements $\phi_{ij}$ and $i,j$ are flavour indices.
The tracelessness condition is enforced by the propagator.
The indices are kept explicitly and the propagator
connecting a field $\phi_{ij}$ to $\phi_{kl}$
is \cite{Sharpe:2001fh}
\begin{equation}
G_{ijkl}(k) = G^c_{ij}(k)\delta_{il}\delta_{jk}-\delta_{ij}\delta_{kl}
G^q_{ik}(k)/n_\mathrm{sea}\,.
\end{equation}
The number of sea quarks $n_\mathrm{sea}$ is what we call $N_F$.
with $G^c_{ij}=i/(p^2-\chi_{ij})$.
The neutral part of the propagator, $G^q_{ik}$,
can contain double poles. In particular for the mass cases we consider:
\ba
G^q_{vv^\prime} &=& i(\chi_1-\chi_4)/(p^2-\chi_1)^2+i/(p^2-\chi_1)\,,
\nonumber\\
G^q_{vs} &=& i/(p^2-\chi_1)\,,
\nonumber\\
G^q_{ss^\prime} &=& i/(p^2-\chi_4)\,.
\ea
$v,s$ denote valence or sea quarks.
The extra parts come from integrating out the $\Phi_0$ \cite{Sharpe:2001fh}
and enforce the condition that $\Phi$ must be traceless.
When constructing the Feynman diagrams, we keep all flavour indices free.
Those that connect to external states get replaced by the value of the
external valence flavour index and the remaining ones are summed over
the sea quark flavours.
In the present calculation, with all sea quarks the same mass, that corresponds
to a factor of $N_F$ for each free flavour index.

For the Majorana, $SU(N_F)\to SO(N_F)$, case we have that
$\Phi=\pi^a X^a$ with $\Phi$ Hermitian, traceless and symmetric.
Hermitian and traceless follow from $SU(N_F)$ and symmetric from
(\ref{commutatorsAdjoint2}). 
Going to the flavour basis for the diagonal elements of $\Phi$ there is no
change w.r.t. the QCD case, but the flavour charged or off-diagonal elements
must be correctly symmetrized. This has to be done both for
the propagator and the connection to the external states, keeping track of
the needed normalization. Afterwards we set the flavour indices connected to
external states to their valence values and sum over the flavours for the
free indices.

For the Dirac adjoint case, $SU(2N_F)\to SO(2N_F)$, case we have that
$\Phi=\pi^a X^a$ with $\Phi$ Hermitian, traceless and
satisfying $X^a J_S = J_S X^{aT}$ and the matrix $\Phi$ is $2N_F\times2N_F$.
Rewriting $\Phi$ with $N_F\times N_F$ matrices leads to the form
\be
\label{matrices1}
\Phi = \left(\begin{array}{cc} \Phi_A & \Phi_C^\dagger\\
                              \Phi_C & \Phi_A^T\end{array}\right)\,,
~~\mathrm{with}~~\lgl A\rgl=0\,,~\Phi_C=\Phi_C^T\,.
\ee
$\Phi_A$ is Hermitian. The elements in $\Phi_A$ correspond
to quark-antiquark states, those in $\Phi_C$ to diquark states.
$\Phi_A$ can be treated exactly as in the QCD case, both the diagonal and
flavour charged or offdiagonal elements, since $\lgl\Phi_A\rgl=0$
replaces $\lgl\Phi\rgl=0$ in the QCD case.
$\Phi_C$ can be treated as offdiagonal or flavour charged propagators
but the needed symmetrizing should be taken care of both for external
states and propagators.
The normalization of all states must be done correctly as well.
After constructing Feynman diagrams with both $\Phi_A$ and $\Phi_C$ degrees
of freedom taken into account, we sum free index lines over the $N_F$ degrees
of freedom, not $2N_F$. The results always agree with the 
calculations done with the previous, Majorana, method.

For the last case, $SU(2N_F)\to Sp(2N_F)$, pseudo-real, we have that
$\Phi=\pi^a X^a$ with $\Phi$ Hermitian, traceless and
satisfying $X^a J_A = J_A X^{aT}$ and the matrix $\Phi$ is $2N_F\times2N_F$.
Rewriting $\Phi$ with $N_F\times N_F$ matrices leads to the form
\be
\label{matrices2}
\Phi = \left(\begin{array}{cc} \Phi_A & \Phi_C^\dagger\\
                              \Phi_C & \Phi_A^T\end{array}\right)\,,
~~\mathrm{with}~~\lgl A\rgl=0\,,~\Phi_C=-\Phi_C^T\,.
\ee
$\Phi_A$ is Hermitian. The elements in $\Phi_A$ correspond
to quark-antiquark states, those in $\Phi_C$ to diquark states.
$\Phi_A$ can be treated exactly as in the QCD case, both the diagonal and
flavour charged or offdiagonal elements, since $\lgl\Phi_A\rgl=0$
replaces $\lgl\Phi\rgl=0$ in the QCD case.
$\Phi_C$ can be treated as offdiagonal or flavour charged propagators
but the needed antisymmetrizing should be taken care of.
The normalization of all states must be done correctly as well.
After constructing Feynman diagrams with both $\Phi_A$ and $\Phi_C$ degrees
of freedom taken into account, we sum free index lines over the $N_F$ degrees
of freedom, not $2N_F$. In this case and the previous we can also
compare calculations with $\Phi_A$ or $\Phi_C$ external states providing a
check on our results.

\section{Analytical results}
\label{analytical}

We have calculated the masses, decay constants and vacuum expectation values
to NNLO  for the QCD-like theories with the symmetry breaking
patterns discussed above.
A number of checks have been performed on the analytical formulas.
The infinite volume unquenched results were obtained earlier
in \cite{Bijnens:2009qm} and we have reproduced those.
The partially quenched and finite volume results in the QCD case are finite.
The partially quenched expressions reduce to the unquenched results
whenever we set the sea mass equal to the valence mass.
In addition we reproduce the known results at NLO for the condensate
\cite{Toublan:1999hi} also for the partially quenched case.
The finite volume expressions have been checked against the known NLO results
and numerically with the earlier known NNLO results, as discussed in
Sect.~\ref{numerical}.

For the real and pseudo-real case we have the additional check that calculating
the mass or decay constant of a quark-anti-quark or a diquark meson
gives the same results. This corresponds to using a field from the
$A$ or the $C$ sector in the matrices (\ref{matrices1},\ref{matrices2}). For the real case
we have the additional check that the results using the Dirac case and
the Majorana case coincide.

The finite volume case is always done for three spatial dimensions of size
L and an infinite temporal volume. In addition we work in the center of mass
system, the momenta are such that the external states have zero spatial
momentum.

The masses are the physical masses as defined as the pole of the
full propagator. We consider here the case where all valence quarks
have the same quark mass $m_1 = \hat m$ and the sea quarks all have the same
mass $m_4 = m_S$. For the unquenced case obviously $m_4=m_1$.
The labeling is similar to those used in three flavour PQChPT
\cite{Bijnens:2004hk,Bijnens:2005ae,Bijnens:2006jv,Bijnens:2015dra}.
In the formulas we use instead the quantities
\be
\chi_1 = 2B_0 m_1,\quad \chi_4 = 2 B_0\chi_4,\quad
\chi_{14}=\frac{1}{2}\left(\chi_1+\chi_4\right)\,.
\ee
These quantities are referred to in \cite{analyticalresults} as m11, m44 and
m14 respectively.

The formulas are given for the cases $SU(N_F)\times SU(N_F)\to SU(N_F)$,
$SU(N_F)\to SO(N_F)$ and $SU(2N_F)\to Sp(2N_F)$.
Note the difference in convention for the second case compared to
\cite{Bijnens:2009qm}. The three cases are referred to in the formulas with
SUN, SON and SPN for the unquenced case and PQSUN,PQSON and PQSPN for
the partially quenched case. In the latter case $N_F$ referes to the number
of sea quarks.

For the mass we consider a meson made of a different quark and anti-quark
or a diquark state with two different quarks.
These are always valence quarks. The physical mass at finite volume is
given by
\be
m_\mathrm{phys}^2 =
\chi_1+ m^{(4)2}+\Delta^V m^{(4)2}+m^{(6)2}+\Delta^V m^{(6)2}\,.
\ee
The superscript $(n)$ labels the order $p^n$ correction and
$\Delta^V$ indicates the finite volume corrections.
In all cases the lowest order mass squared is given by $\chi_1$.
A further break up is done for the LEC dependent parts via the
$L_i^r$ (NLO) and $K_i^r$ (NNLO) and the remainder via
\ba
m^{(4)2} &=& m^{L(4)2}+m^{R(4)2}
\nonumber\\
m^{(6)2} &=& m^{K(6)2}+m^{L(6)2}+m^{R(6)2}
\nonumber\\
\Delta^V m^{(6)2} &=& \Delta^V  m^{L(6)2}+\Delta^V m^{R(6)2}
\ea
All quantities are given explicitly in \cite{analyticalresults}.

The decay constant $F_\mathrm{phys}$ for the same mesons as above is
expanded w.r.t. the lowest order as
\be
F_\mathrm{phys} =
F_{LO}\left(1+ F^{(4)}+\Delta^V F^{(4)}+F^{(6)2}+\Delta^V F^{(6)}\right)\,,
\ee
with a similar split
\ba
F^{(4)} &=& F^{L(4)}+F^{R(4)}
\nonumber\\
F^{(6)} &=& F^{K(6)}+F^{L(6)}+F^{R(6)}
\nonumber\\
\Delta^V F^{(6)} &=&\Delta^V  F^{L(6)}+\Delta^V F^{R(6)}
\ea
All quantities are given explicitly in \cite{analyticalresults}.

The vacuum expectation value is expanded in exactly the same way
\be
v_\mathrm{phys} =
v_{LO}\left(1+ v^{(4)}+\Delta^V v^{(4)}+v^{(6)2}+\Delta^V v^{(6)}\right)\,,
\ee
with a similar split
\ba
v^{(4)} &=& v^{L(4)}+v^{R(4)}
\nonumber\\
v^{(6)} &=& v^{K(6)}+v^{L(4)}+v^{R(4)}
\nonumber\\
\Delta^V v^{(6)} &=&\Delta^V  v^{L(6)}+\Delta^V v^{R(6)}
\ea
All quantities are given explicitly in \cite{analyticalresults}.

The quantities with $K$ for the SON and SPN case have been set to zero.
They are polynomials up to the needed degree in $\chi_1$ and $\chi_4$,
with an overall factor of $\chi_1$ for the mass.

The decay constant and the vacuum expectation value were defined implicitly
in \cite{Bijnens:2009qm} using a generator $X^a$ in the axial current
normalized to one and an element in $\hat\mathcal{M}$ normalized to one.
The consequence was that in \cite{Bijnens:2009qm} $F_{LO}=F$ and
$v_{LO}=-B_0F^2$ for all cases.
This is not exactly what was done in earlier work leading to differences
in factors of $2$ and $\sqrt{2}$. Below we explicitly specify all definitions
in terms of the quark fields.

\paragraph{QCD or complex representation}

If we label the first Dirac (valence) quark by 1 and the second by 2
the decay constant and vacuum expectation value are defined as
\ba
\lgl 0|\overline q_1 \gamma_\mu\gamma_5 q_2|M(p)\rgl &=&
i\sqrt{2} F_\mathrm{phys} p_\mu
\nonumber\\
\lgl \overline q_1 q_1\rgl&=&
\lgl \overline q_{L1}q_{R1}+ \overline q_{R1}q_{L1}\rgl
= v_\mathrm{phys}
\ea
$M$ denotes a meson of that quark content with momentum $p$.

The resulting lowest orders are
\be
F_{LO} = F\,\quad v_{LO} = -B_0 F^2\,.
\ee

\paragraph{Adjoint or real representation}

Here we have to be careful how we define the physical decay constant.
We can choose to do using generators normalized to one using Dirac Fermions
or generators normalized to one using the $\hat q_i$ elements.

With a Dirac fermion definition,
the first Dirac (valence) quark labeled by 1 and the second by 2,
the definitions are
\ba
\lgl 0|\overline q_1 \gamma_\mu\gamma_5 q_2|M(p)\rgl &=&
i\sqrt{2} F_\mathrm{phys} p_\mu
\nonumber\\
\lgl \overline q_1 q_1\rgl&=&
\lgl \overline q_{L1}q_{R1}+ \overline q_{R1}q_{L1}\rgl
= v_\mathrm{phys}
\ea
$M$ denotes a meson of that quark content with momentum $p$.
The resulting lowest orders are
\be
F_{LO} = \sqrt{2}F\,\quad v_{LO} = -2B_0 F^2\,.
\ee

If we instead choose to use the Majorana case, the natural definition
of the decay constant and vacuum expectation value with the first (valence)
Majorana fermion
labeled as 1 and the second as 2 via
\ba
\frac{1}{2\sqrt{2}}\lgl 0|
 \hat q_1^* \overline\sigma_\mu \hat q_2
+\hat q_2^* \overline\sigma_\mu \hat q_1|M(p)\rgl &=&
i\sqrt{2} F_\mathrm{phys} p_\mu
\nonumber\\
\frac{1}{2}\lgl \hat q_{1}\sigma^2\hat q_{1}
              + \hat q_{1}^*\sigma^2\hat q_{1}^*\rgl
&=& v_\mathrm{phys}
\ea
The resulting lowest orders are
\be
F_{LO} = F\,\quad v_{LO} = -B_0 F^2\,.
\ee

\paragraph{$N_c=2$ or pseudo-real representation}

Here we again need to be careful how we define the physical decay constant.
We can choose to do using generators normalized to one using
the original Dirac Fermions
or generators normalized to one using the $\hat q_i$ elements.

With a Dirac fermion definition,
the first Dirac (valence) quark labeled by 1 and the second by 2,
the definitions are
\ba
\lgl 0|\overline q_1 \gamma_\mu\gamma_5 q_2|M(p)\rgl &=&
i\sqrt{2} F_\mathrm{phys} p_\mu
\nonumber\\
\lgl \overline q_1 q_1\rgl&=&
\lgl \overline q_{L1}q_{R1}+ \overline q_{R1}q_{L1}\rgl
= v_\mathrm{phys}
\ea
$M$ denotes a meson of that quark content with momentum $p$.
The resulting lowest orders are
\be
F_{LO} = \sqrt{2}F\,\quad v_{LO} = -2B_0 F^2\,.
\ee

In terms of the $\hat q_i$
the definitions are
\ba
\lgl 0|\overline{\hat q}_1 \gamma_\mu\overline{\hat q}_2
      +\overline{\hat q}_{1+N_F} \gamma_\mu\overline{\hat q}_{2+N_F}
       |M(p)\rgl &=&
i\sqrt{2} F_\mathrm{phys} p_\mu
\nonumber\\
\frac{1}{2}\lgl
 \overline{\hat q}_{1+N_F,a}\epsilon_{ab}C\overline{\hat q}_{1,b}
-\overline{\hat q}_{1,a}\epsilon_{ab}C\overline{\hat q}_{1+N_F,b}
-{\hat q}_{1+N_F,a}\epsilon_{ab}C{\hat q}_{1,b}
+{\hat q}_{1,a}\epsilon_{ab}C{\hat q}_{1+N_F,b}
&=& v_\mathrm{phys}\,.
\ea

\section{Numerical examples and checks}
\label{numerical}

The main aim of this work is to provide the lattice work with the formulas and
programs needed to do the extrapolation to zero mass. We therefore only present
some representative numerical results.
The numerical programs are included in the latest version of \textsc{CHIRON},
\cite{Bijnens:2014gsa,chiron}.

For the numbers presented we always use $\chi_1=0.14^2$~GeV$^2$,
if not varied explicitly, and $F=0.0877$~GeV as well as a subtraction scale
$\mu=0.77$~GeV. The length $L$ for the finite volume has been chosen such that
$L\times0.14$~GeV=3 or $L\approx4.2$~fm.

The LECs at NLO we choose to be those of the recent determination
of \cite{Bijnens:2014lea} with the extra LEC $L_0^r=0$.
The NNLO constants we have always put to zero.

A number of numerical checks for the QCD case have been done.
The unquenched infinite volume results for three flavours agree with the
three flavour results of \cite{Amoros:1999dp,Amoros:2000mc}.
The partially quenched results for masses and decay constants at infinite
volume agree with the case $d_\mathrm{sea}=1,d_\mathrm{val}=1$
of \cite{Bijnens:2004hk,Bijnens:2005ae,Bijnens:2006jv}.
The unquenched results for masses and decay constants at finite
volume agree with \cite{Bijnens:2014dea}.
The partially quenched results for masses and decay constants at finite
volume agree with the case $d_\mathrm{sea}=1,d_\mathrm{val}=1$
of \cite{Bijnens:2015dra} and finially the unquenched finite volume results
for the vacuum expectation value agree with the results
of \cite{Bijnens:2006ve}.

\begin{figure}
\begin{center}
\includegraphics[width=0.43\textwidth]{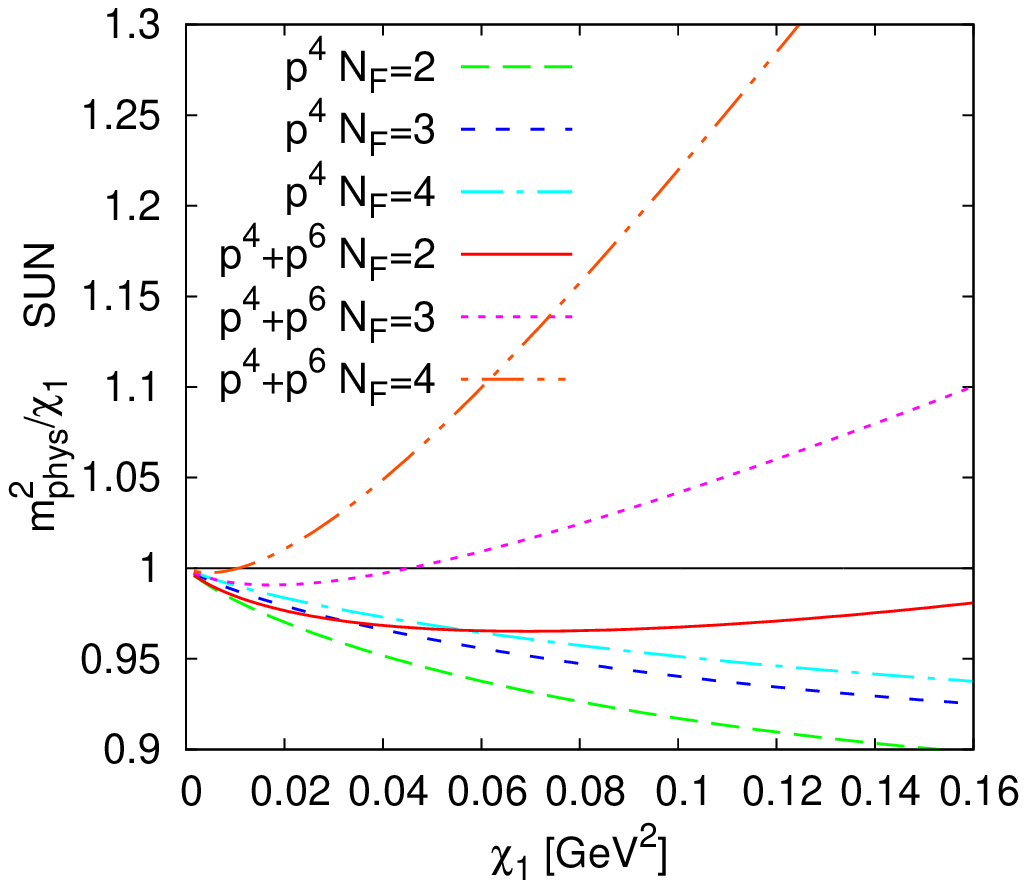}
\includegraphics[width=0.43\textwidth]{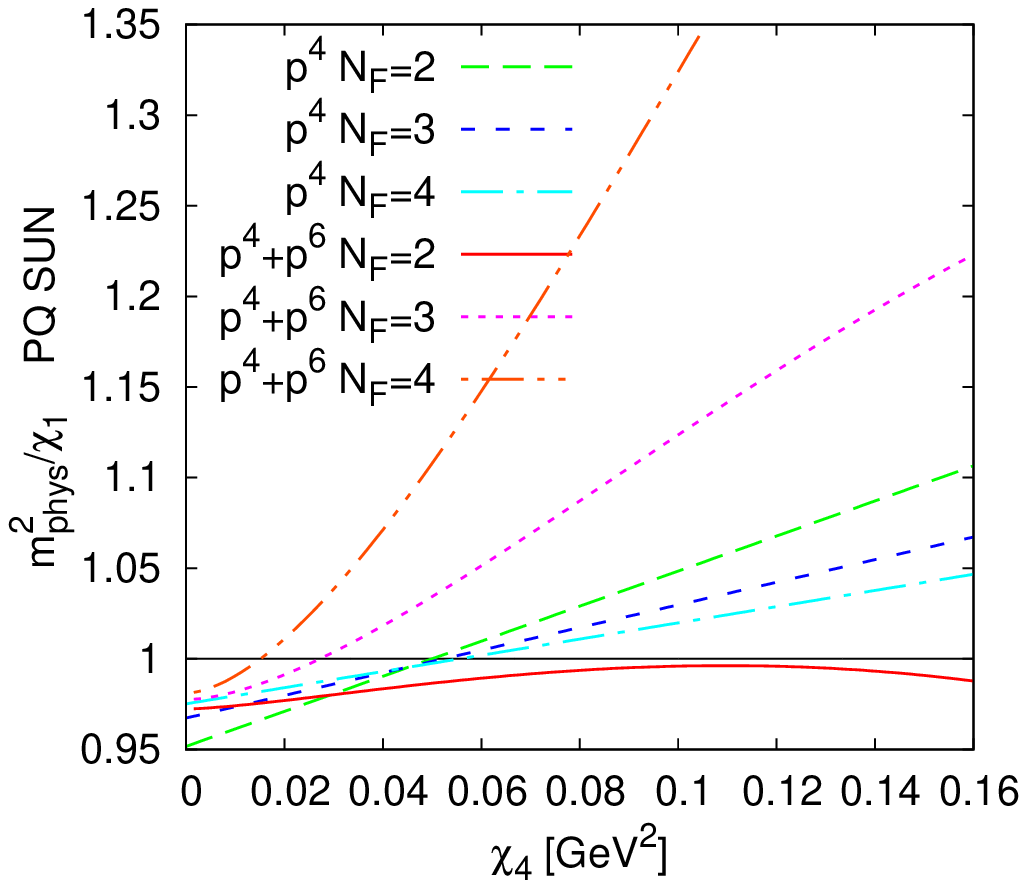}\\[-3mm]
\includegraphics[width=0.43\textwidth]{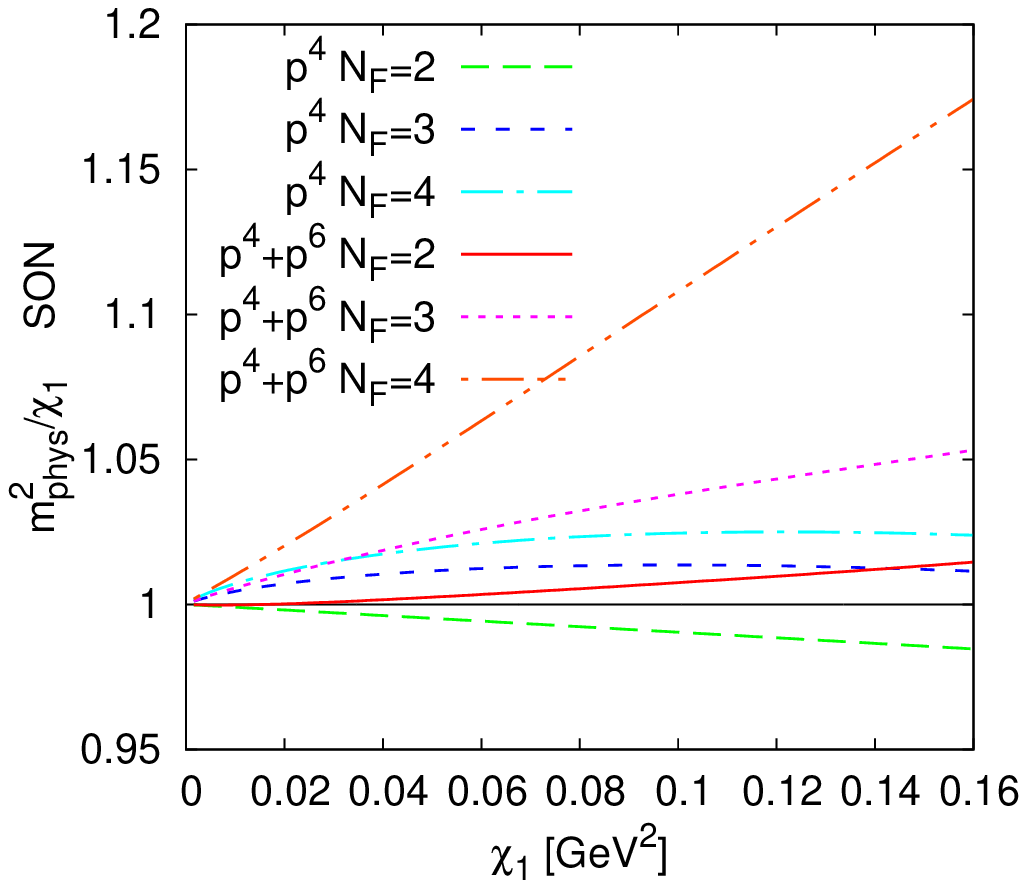}
\includegraphics[width=0.43\textwidth]{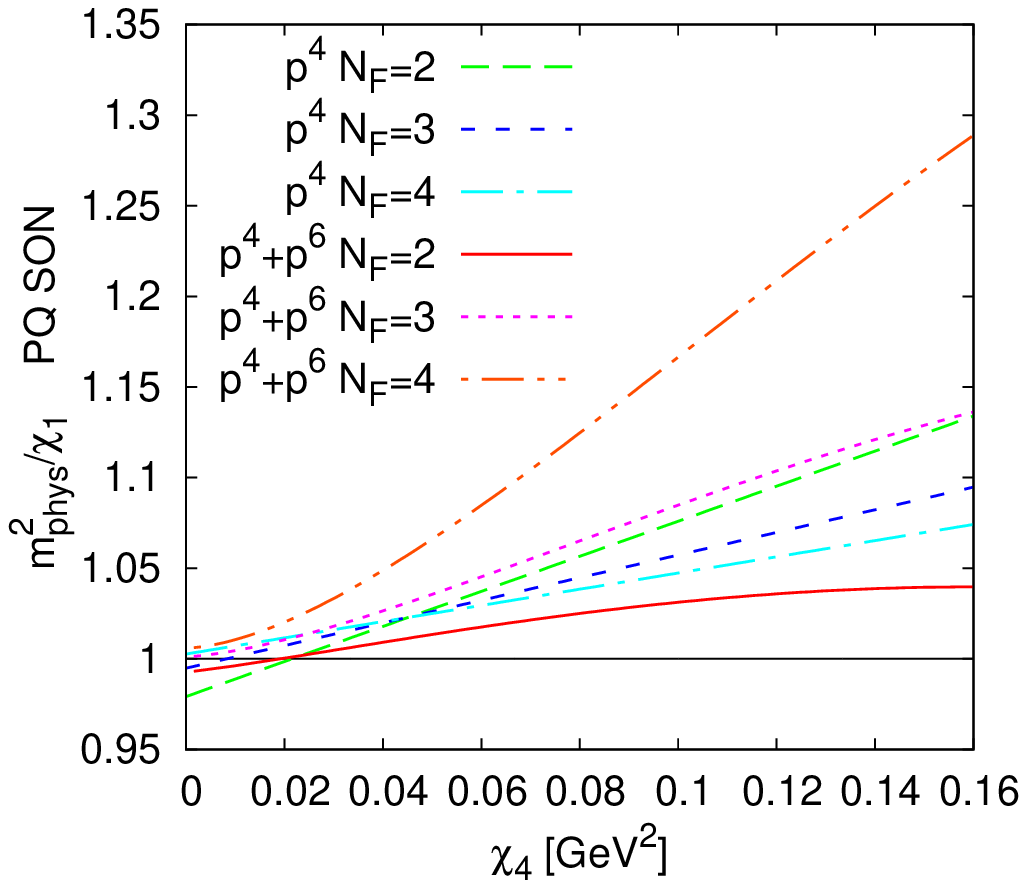}\\[-3mm]
\includegraphics[width=0.43\textwidth]{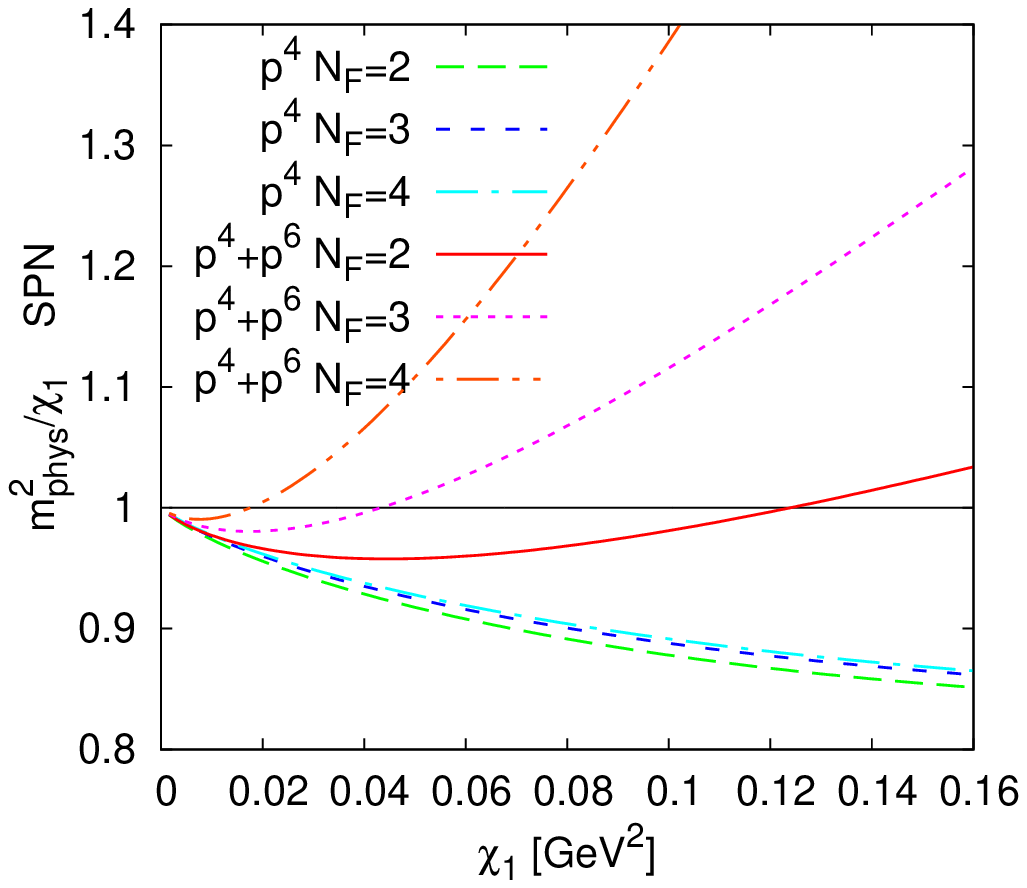}
\includegraphics[width=0.43\textwidth]{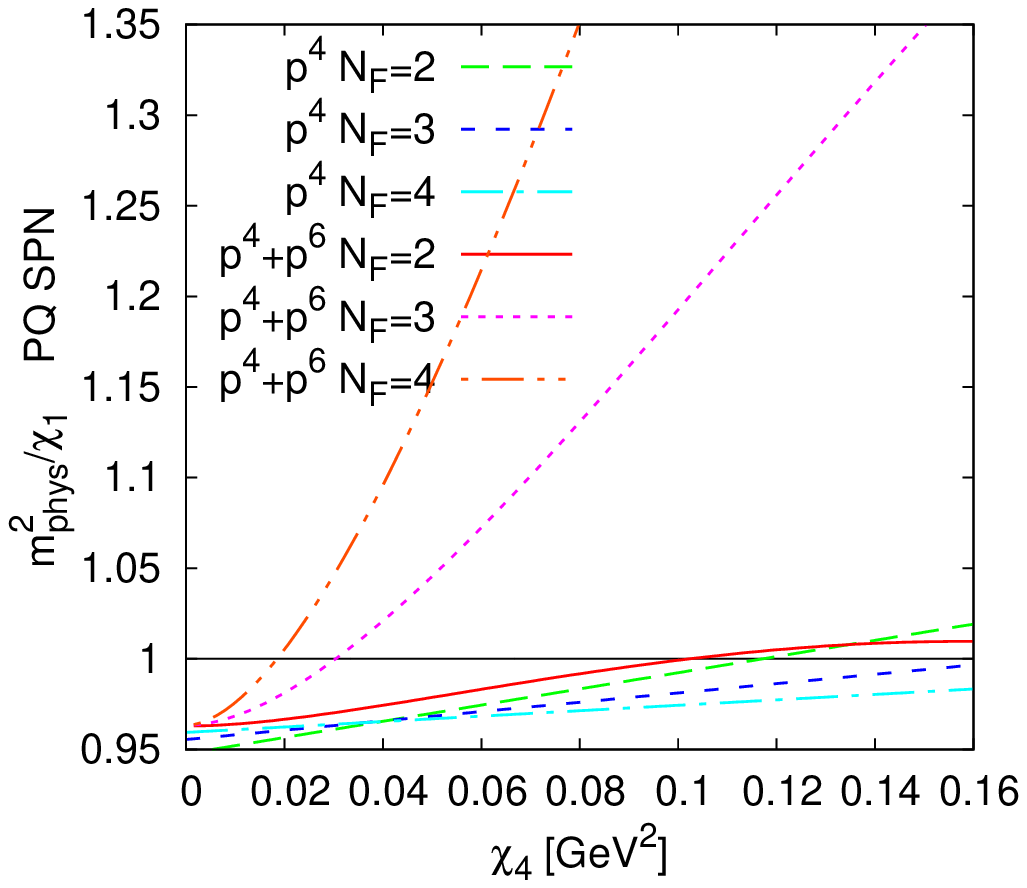}
\end{center}
\caption{\label{figmassIV}
The physical mass squared divided by the lowest order mass squared
for the unquenched (left) as a function of $\chi_1$
and the partially quenched case (right) as a function of $\chi_4$
with $\chi_1=0.14^2$~GeV$^2$. Other input as in the text.
Shown are the NLO ($p^4$) and NNLO ($p^4+p^6$)
results for three values of $N_F$.
Top line: $SU(N_F)\times SU(N_F)\to SU(N_F)$.
Middle line: $SU(N_F)\to SO(N_F)$.
Bottom line: $SU(2N_F)\to Sp(2N_F)$.
}
\end{figure}

In Fig.~\ref{figmassIV} we show the mass squared for the infinite volume
for all cases we have considered for three values of $N_F$.
In general, as was already noticed in \cite{Bijnens:2009qm}
the corrections are larger for the larger values of $N_F$.
The corrections are also larger for the $SU(2N_F)\to Sp(2N_F)$ case
since this correspond to a twice as large number of fermions as the other cases.
The partially quenched results shown in the right column are at a fixed value
of $\chi_1$. That explains why the corrections do not vanish for $\chi_4=0$.

\begin{figure}
\begin{center}
\includegraphics[width=0.43\textwidth]{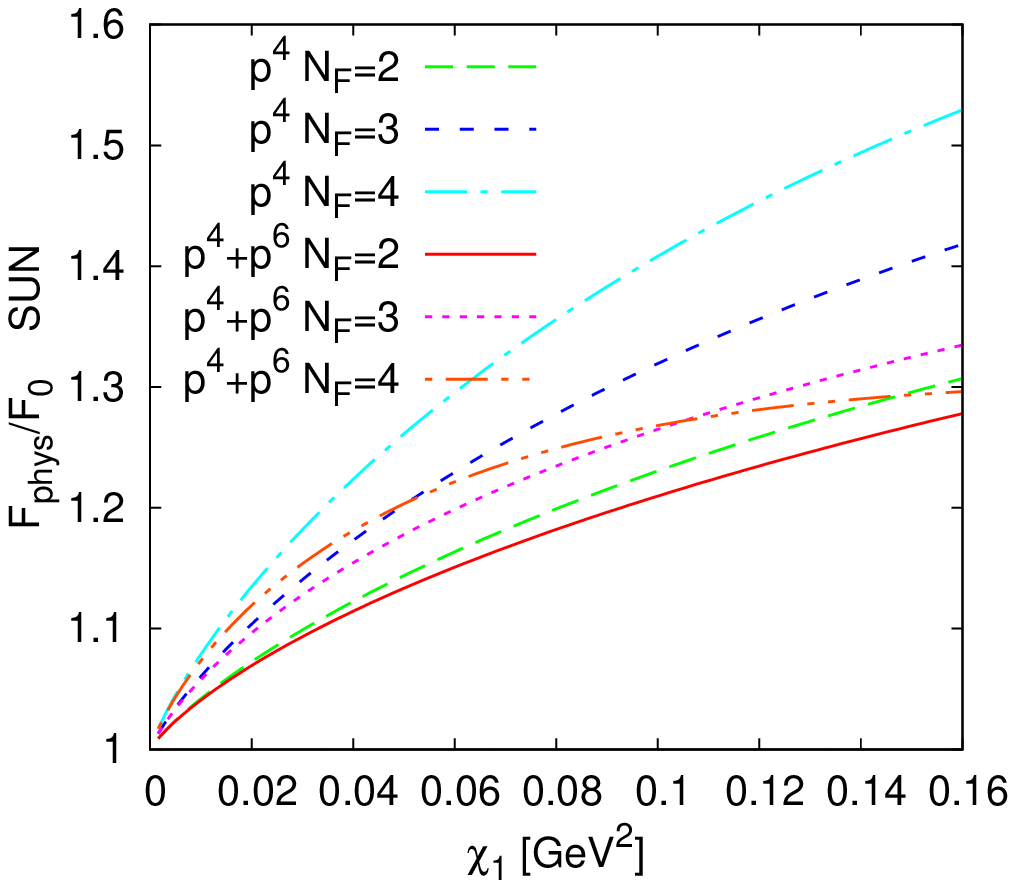}
\includegraphics[width=0.43\textwidth]{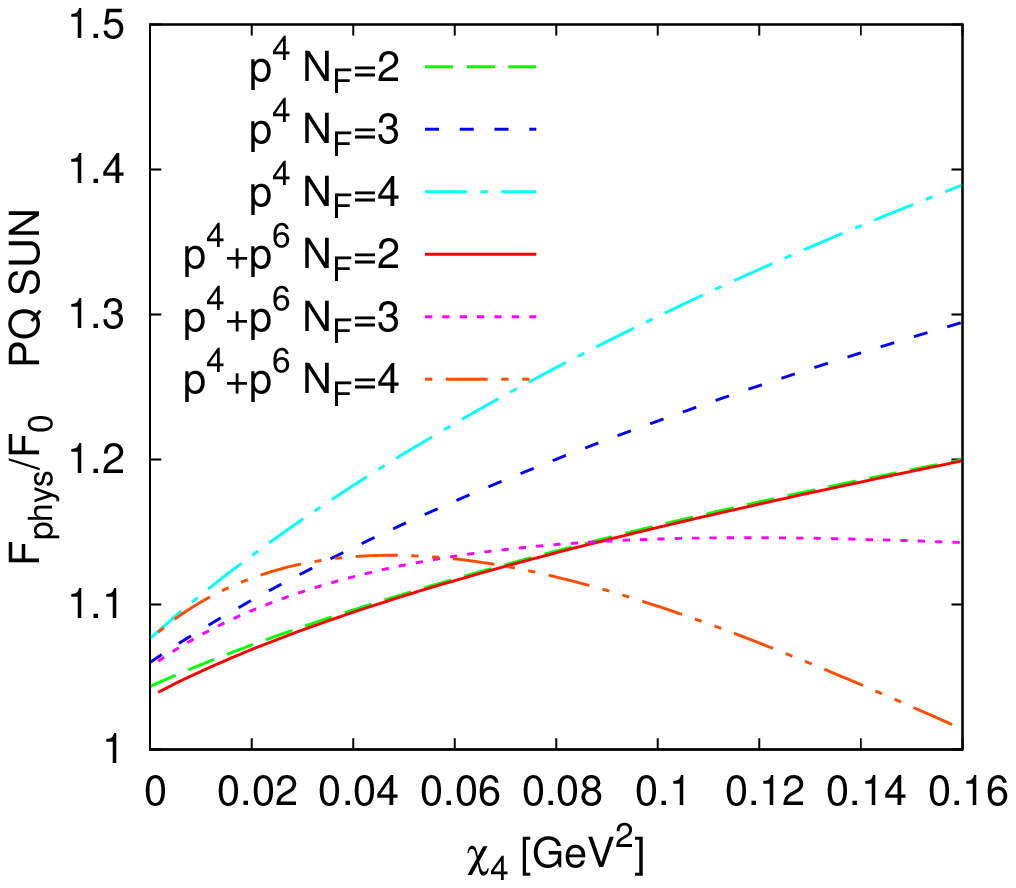}\\[-3mm]
\includegraphics[width=0.43\textwidth]{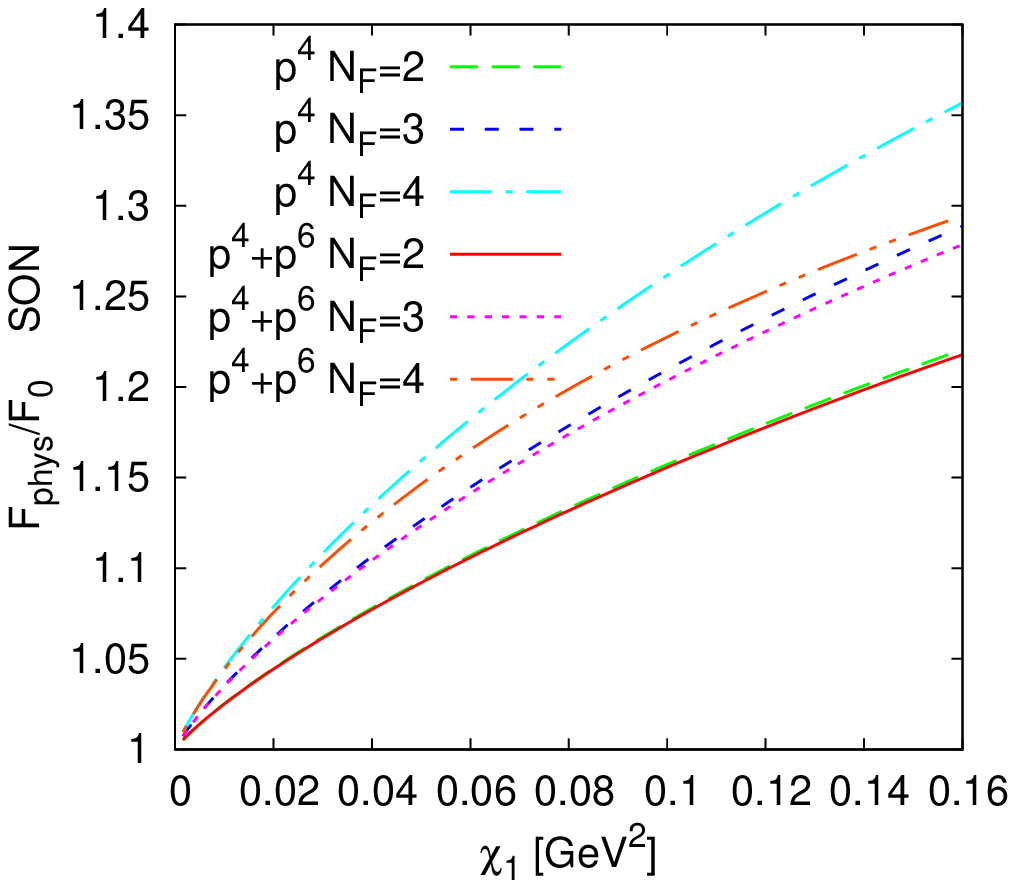}
\includegraphics[width=0.43\textwidth]{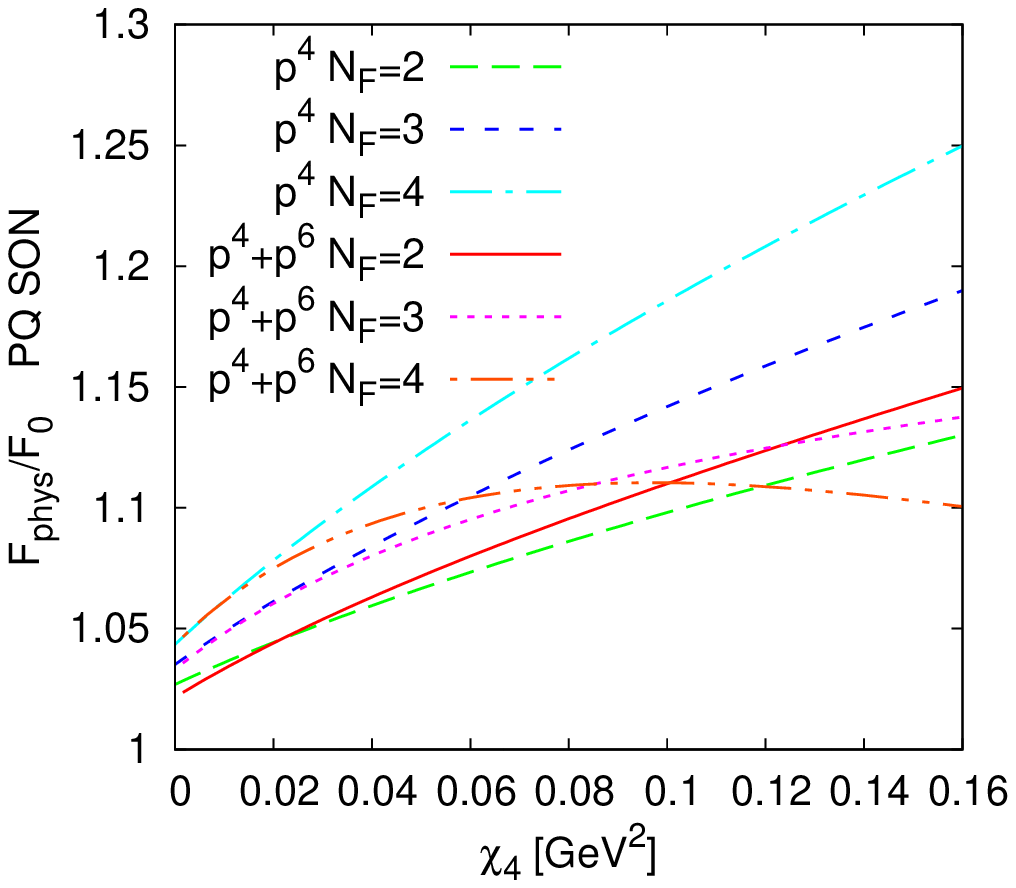}\\[-3mm]
\includegraphics[width=0.43\textwidth]{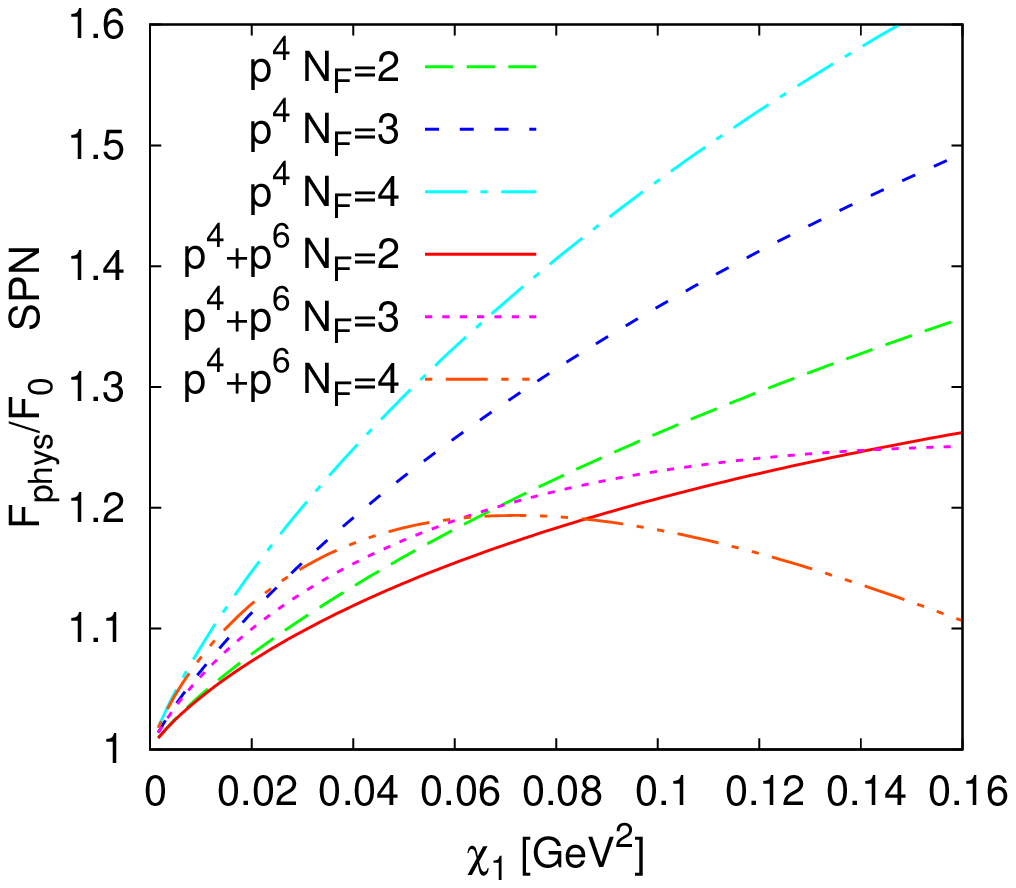}
\includegraphics[width=0.43\textwidth]{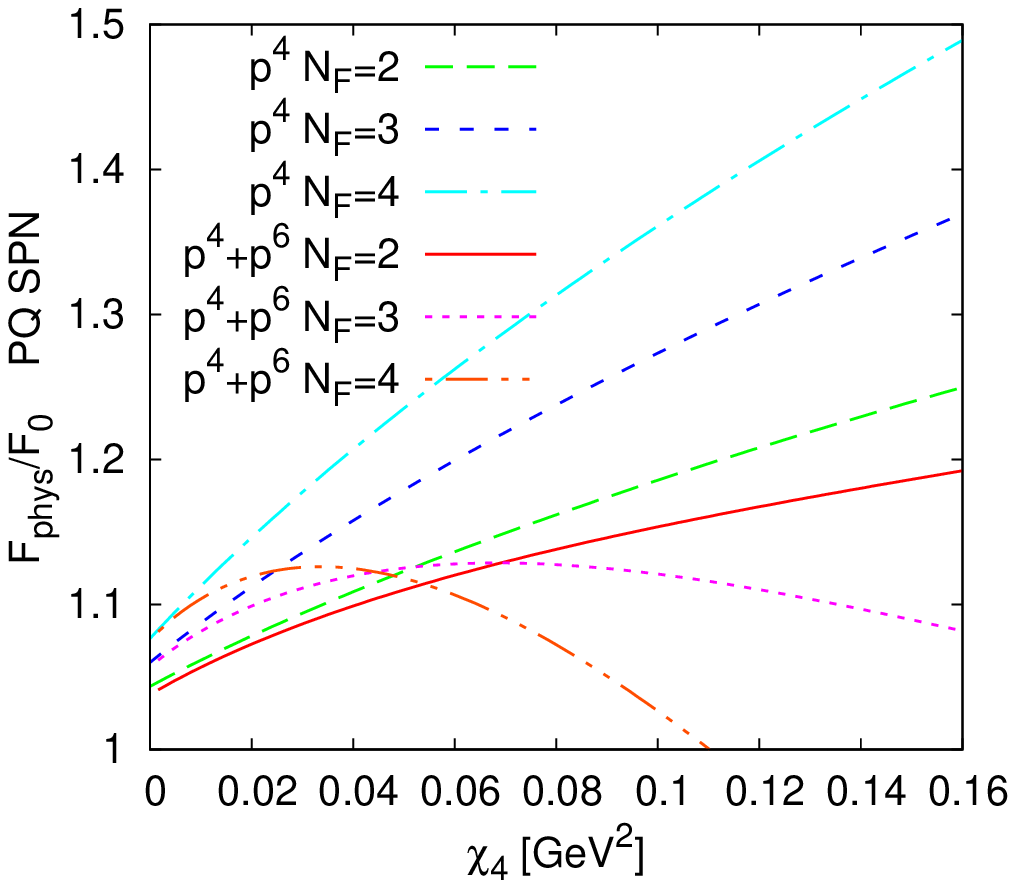}
\end{center}
\caption{\label{figdecayIV}
The decay constant divided by the lowest order value $F_0=F_{LO}$
for the unquenched (left) as a function of $\chi_1$
and the partially quenched case (right) as a function of $\chi_4$
with $\chi_1=0.14^2$~GeV$^2$. Other input as in the text.
Shown are the NLO ($p^4$) and NNLO ($p^4+p^6$)
results for three values of $N_F$.
Top line: $SU(N_F)\times SU(N_F)\to SU(N_F)$.
Middle line: $SU(N_F)\to SO(N_F)$.
Bottom line: $SU(2N_F)\to Sp(2N_F)$.
}
\end{figure}

\begin{figure}
\begin{center}
\includegraphics[width=0.43\textwidth]{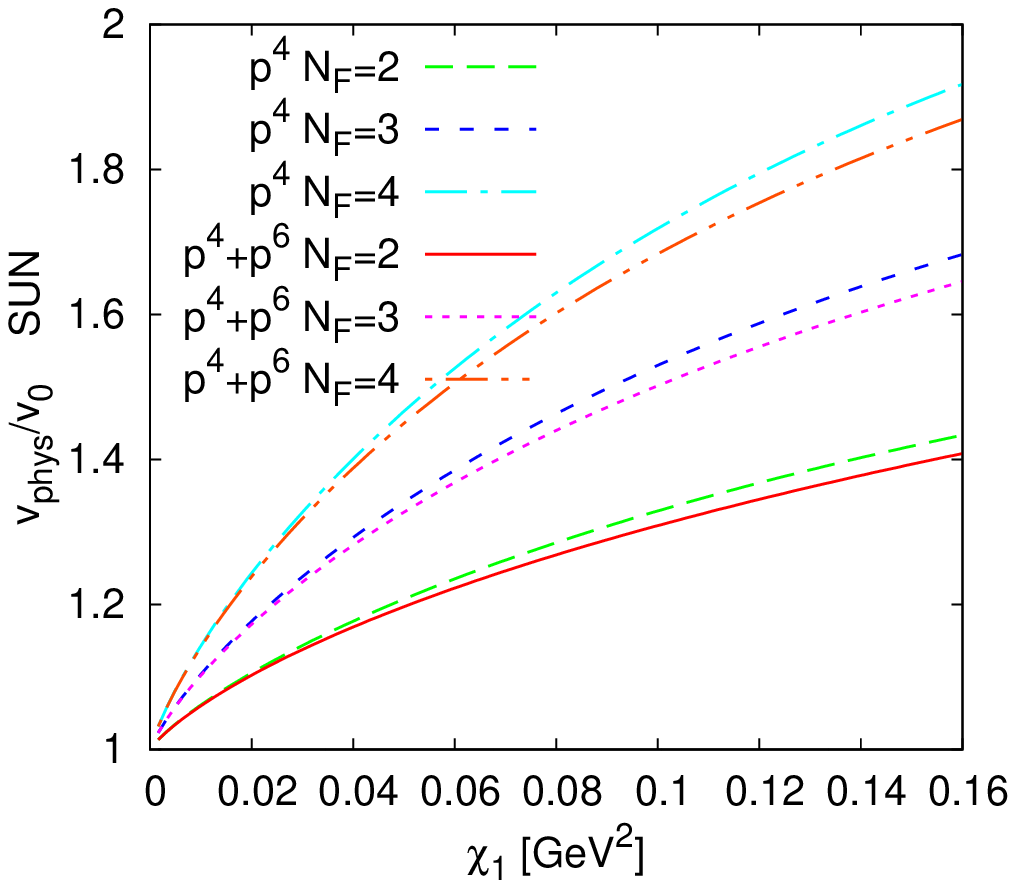}
\includegraphics[width=0.43\textwidth]{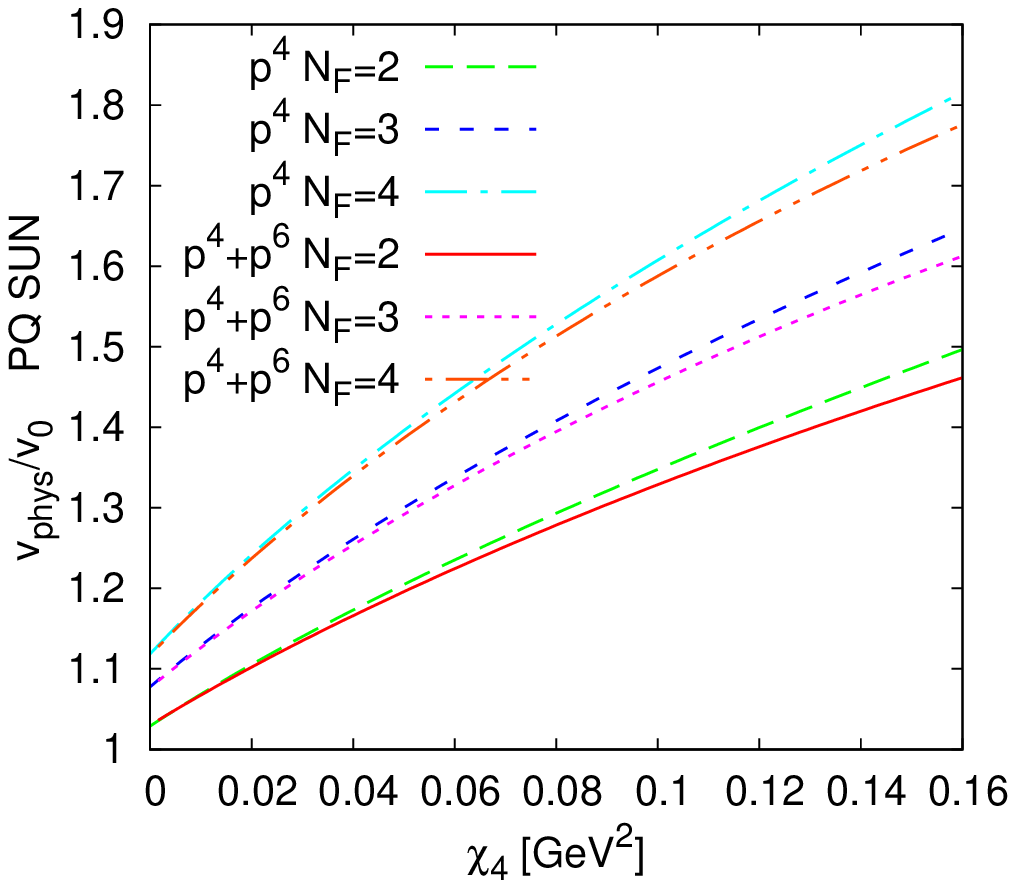}\\[-3mm]
\includegraphics[width=0.43\textwidth]{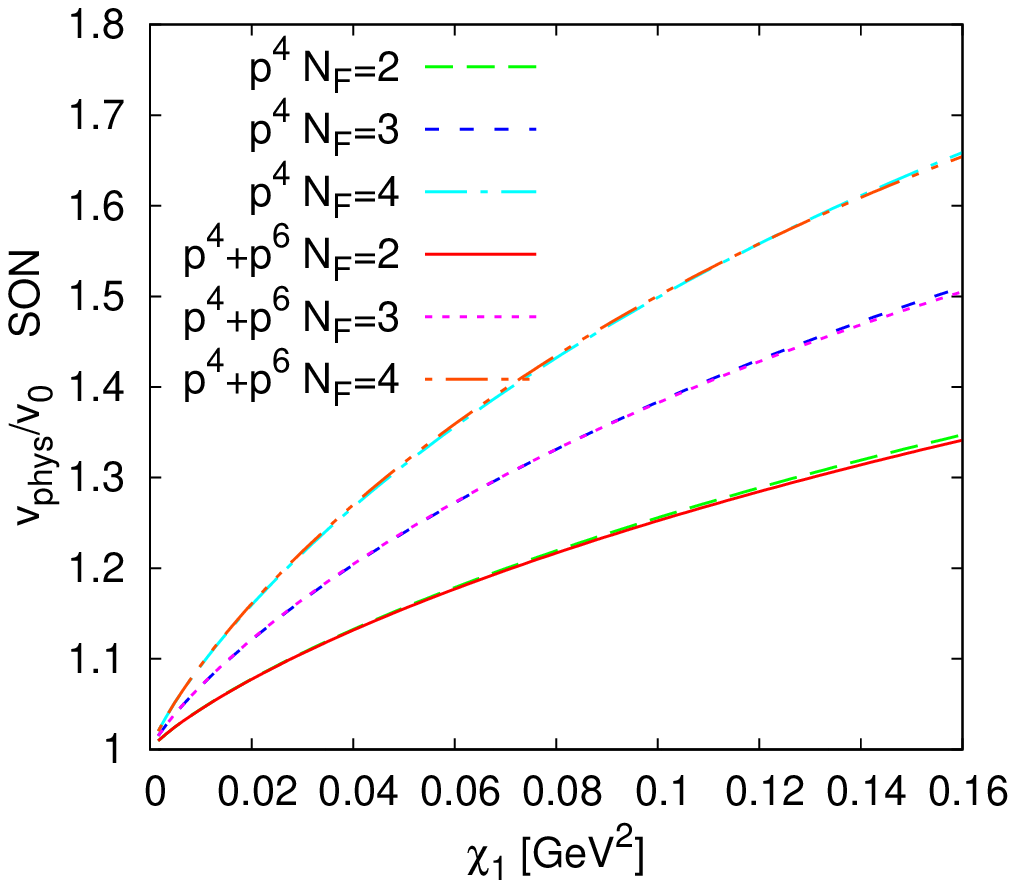}
\includegraphics[width=0.43\textwidth]{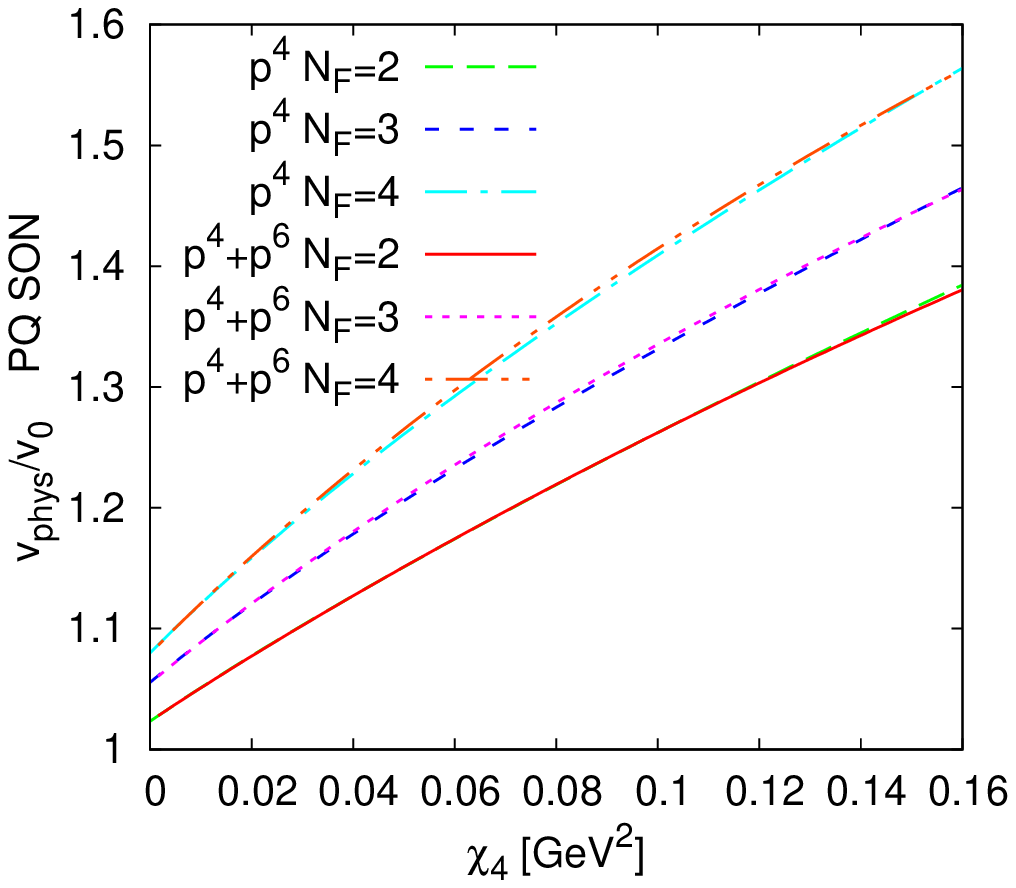}\\[-3mm]
\includegraphics[width=0.43\textwidth]{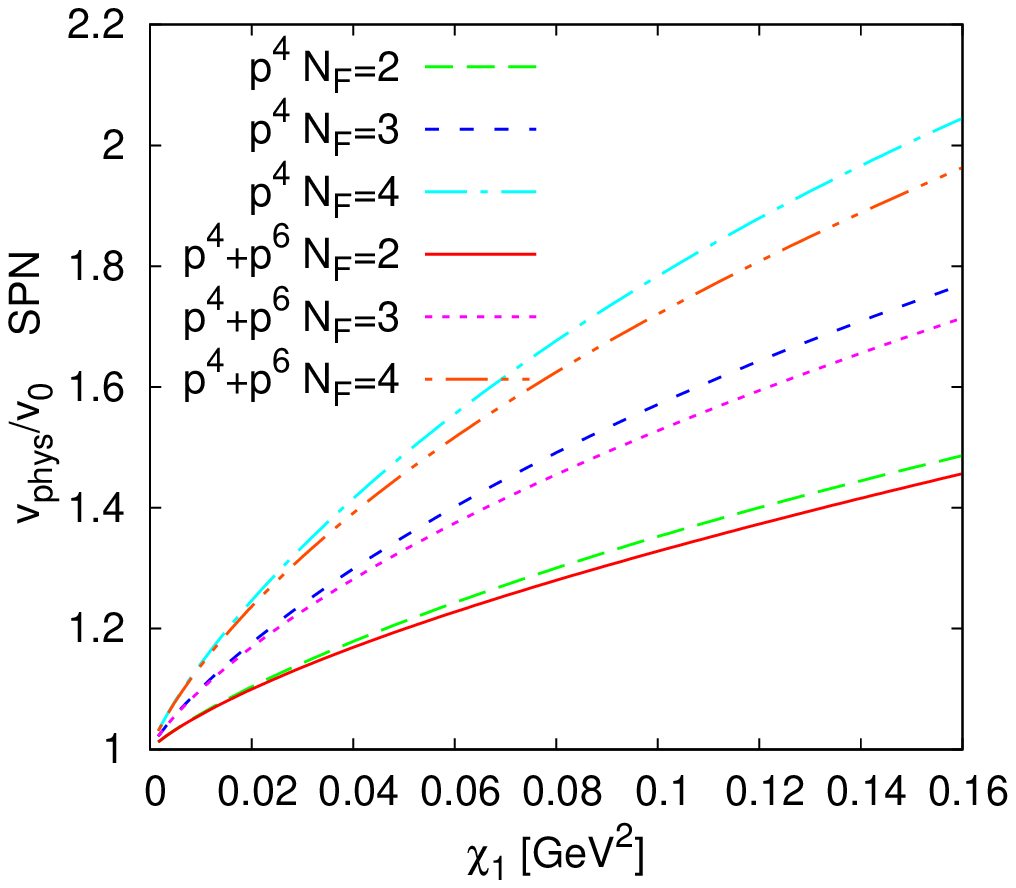}
\includegraphics[width=0.43\textwidth]{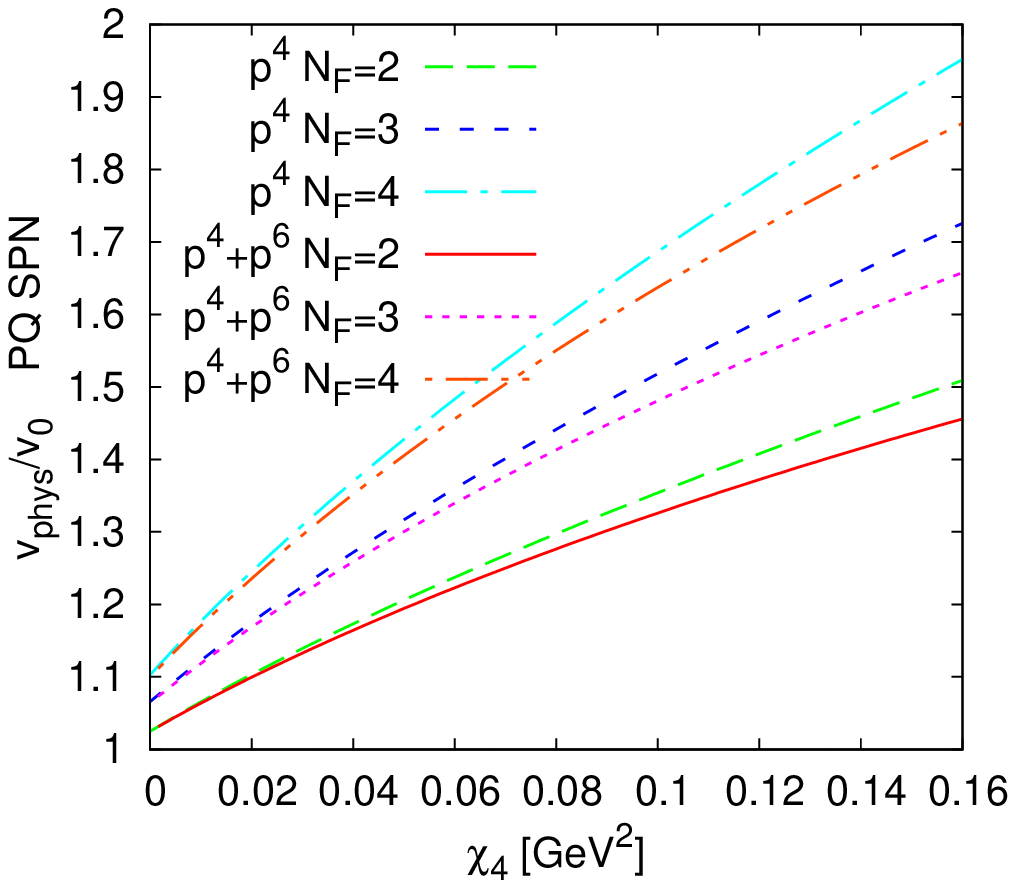}
\end{center}
\caption{\label{figvevIV}
The vacuum expectation value divided by the lowest order value $v_0=v_{LO}$
for the unquenched (left) as a function of $\chi_1$
and the partially quenched case (right) as a function of $\chi_4$
with $\chi_1=0.14^2$~GeV$^2$. Other input as in the text.
Shown are the NLO ($p^4$) and NNLO ($p^4+p^6$)
results for three values of $N_F$.
Top line: $SU(N_F)\times SU(N_F)\to SU(N_F)$.
Middle line: $SU(N_F)\to SO(N_F)$.
Bottom line: $SU(2N_F)\to Sp(2N_F)$.
}
\end{figure}

The same types of results are shown for the decay constant in
Fig.~\ref{figdecayIV}. The corrections are somewhat larger than for the masses
but the convergence is typically somewhat better.
The corrections for the vacuum expectation value shown in Fig.~\ref{figvevIV}
are typically larger but with again a reasonable convergence from
NLO to NNLO.

\begin{figure}
\begin{center}
\includegraphics[width=0.43\textwidth]{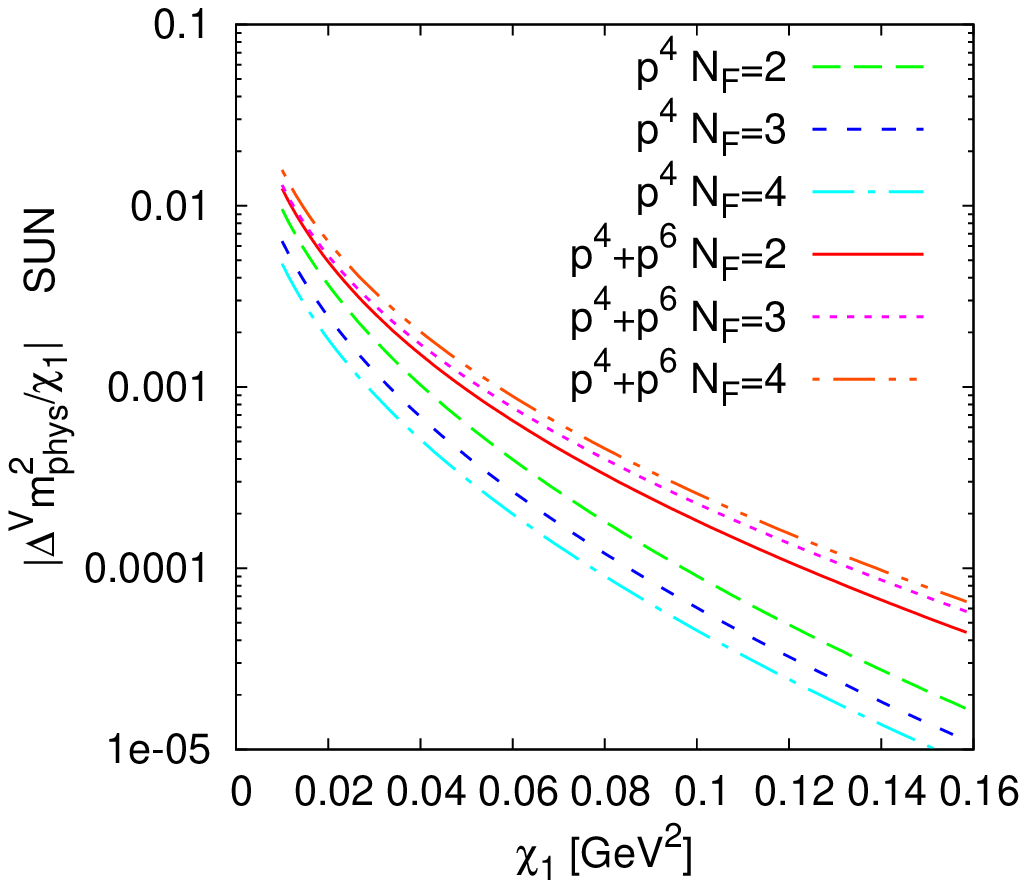}
\includegraphics[width=0.43\textwidth]{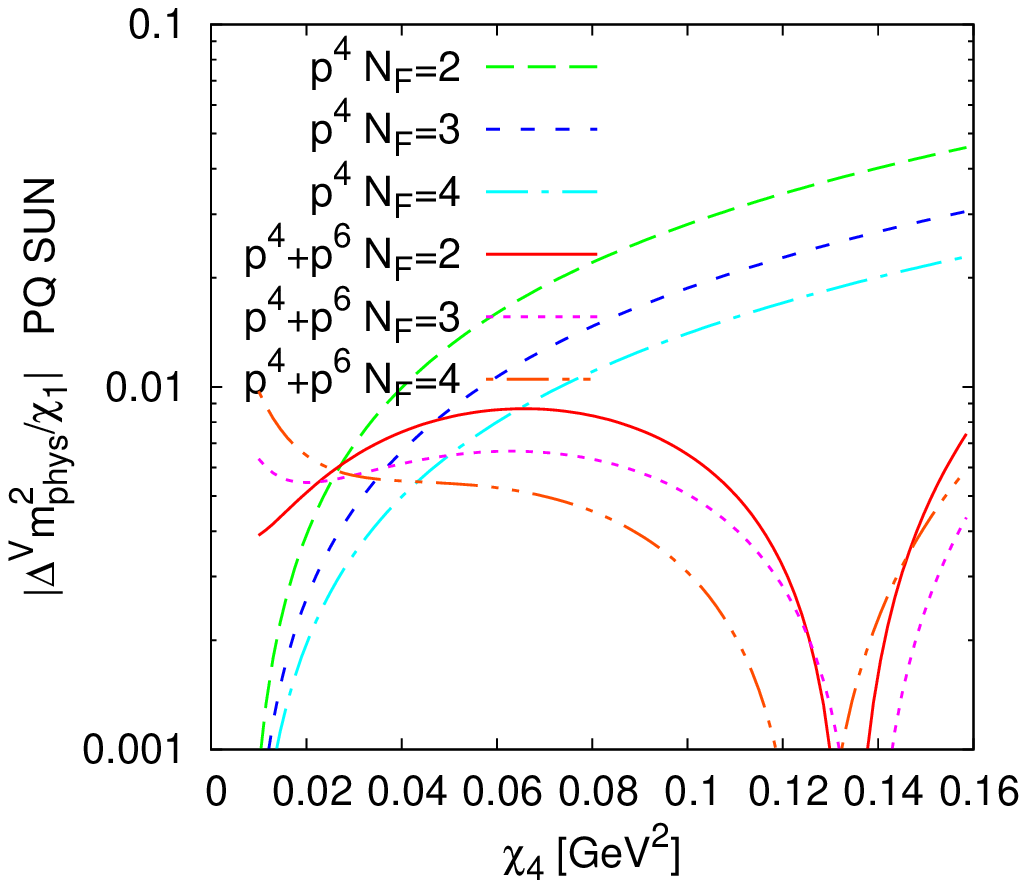}\\[-3mm]
\includegraphics[width=0.43\textwidth]{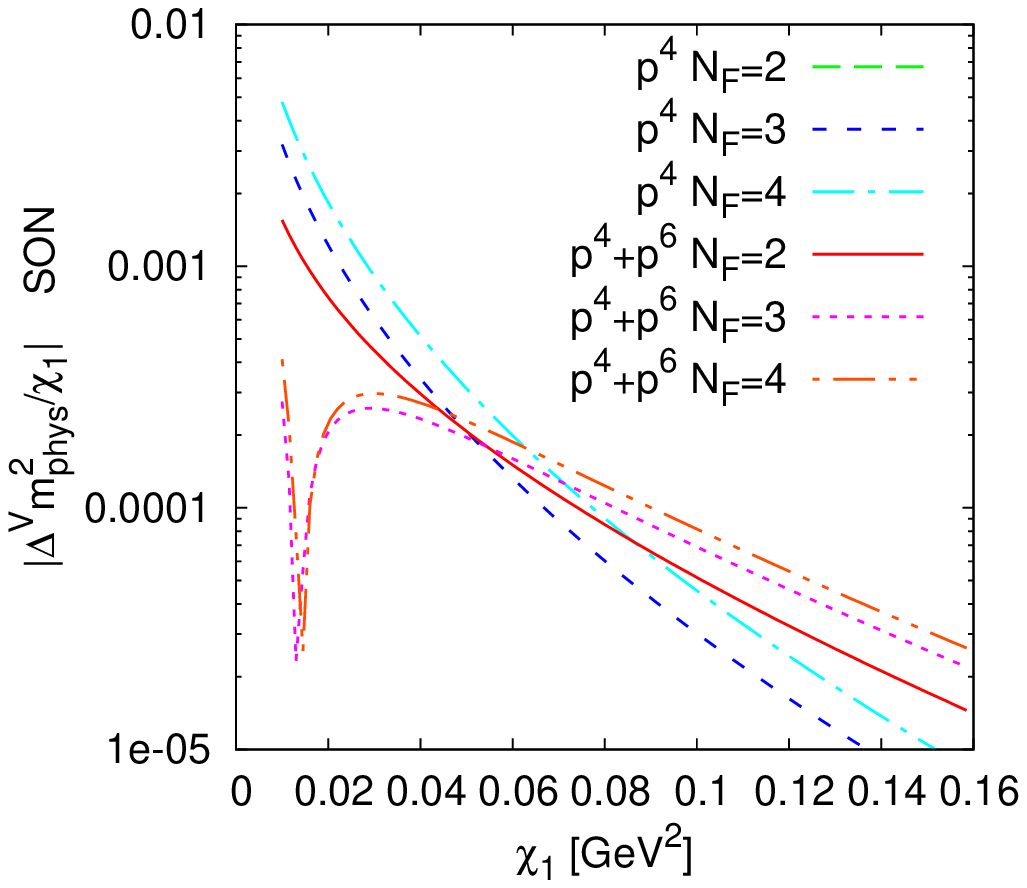}
\includegraphics[width=0.43\textwidth]{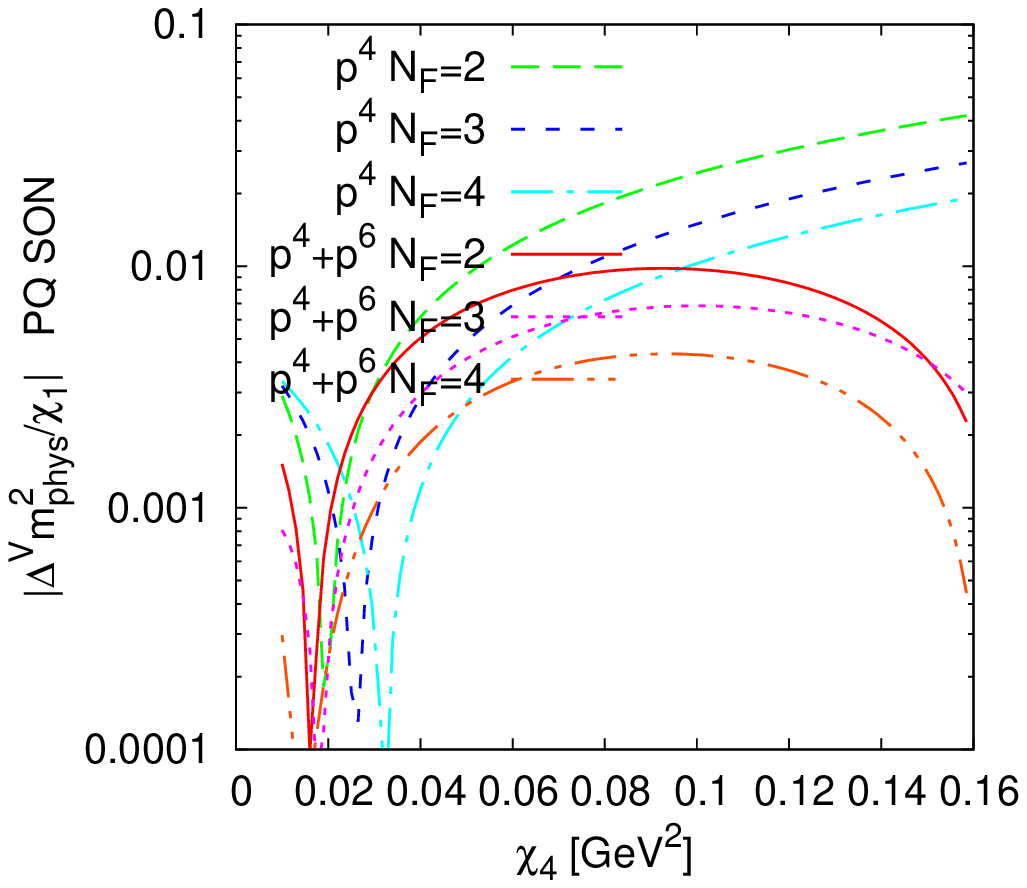}\\[-3mm]
\includegraphics[width=0.43\textwidth]{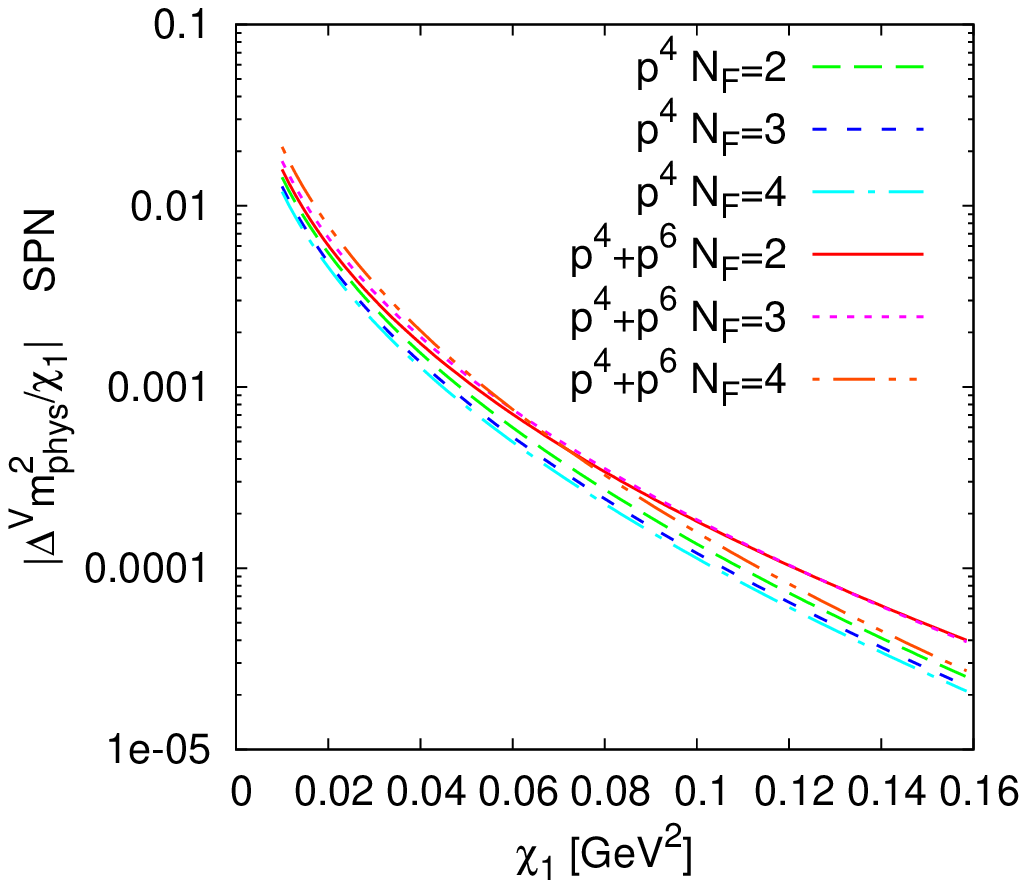}
\includegraphics[width=0.43\textwidth]{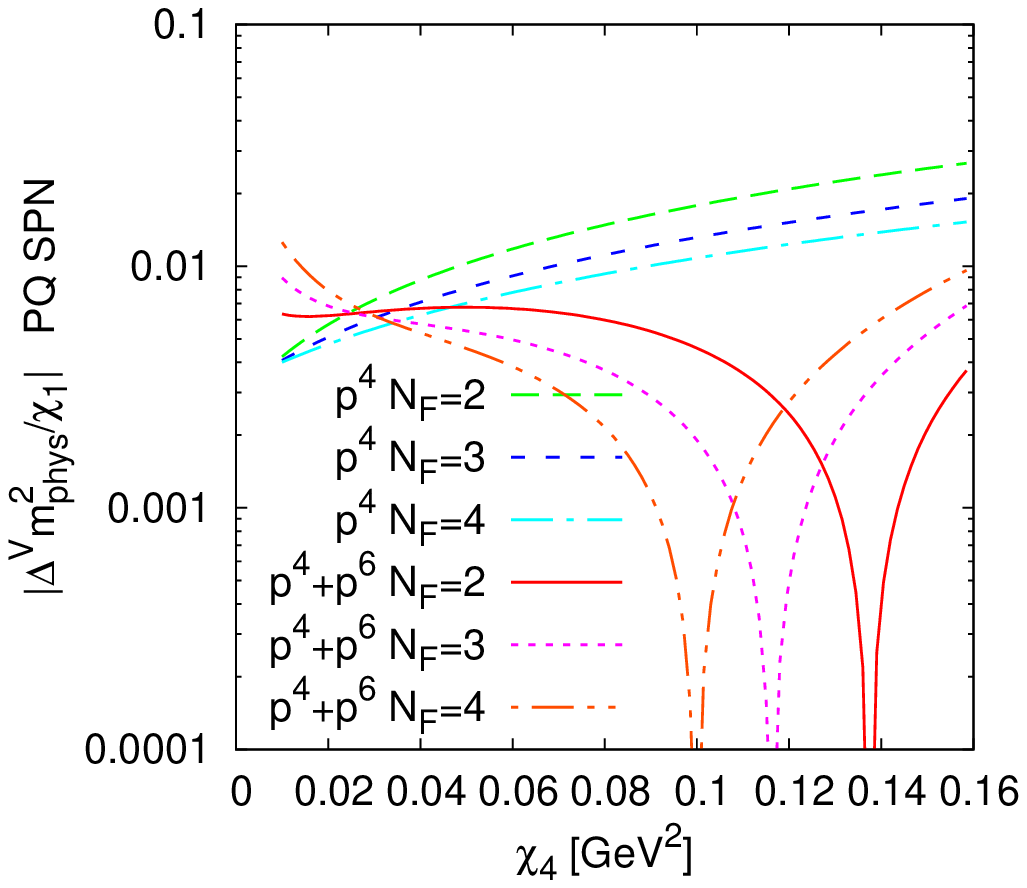}
\end{center}
\caption{\label{figmassFV}
The absolute value of
the finite volume correction to the physical mass squared divided by the
lowest order mass squared
for the unquenched (left) as a function of $\chi_1$
and the partially quenched case (right) as a function of $\chi_4$
with $\chi_1=0.14^2$~GeV$^2$. 
Shown are the NLO ($p^4$) and NNLO ($p^4+p^6$)
results for three values of $N_F$.
Top line: $SU(N_F)\times SU(N_F)\to SU(N_F)$.
Middle line: $SU(N_F)\to SO(N_F)$.
Bottom line: $SU(2N_F)\to Sp(2N_F)$.
}
\end{figure}

\begin{figure}
\begin{center}
\includegraphics[width=0.43\textwidth]{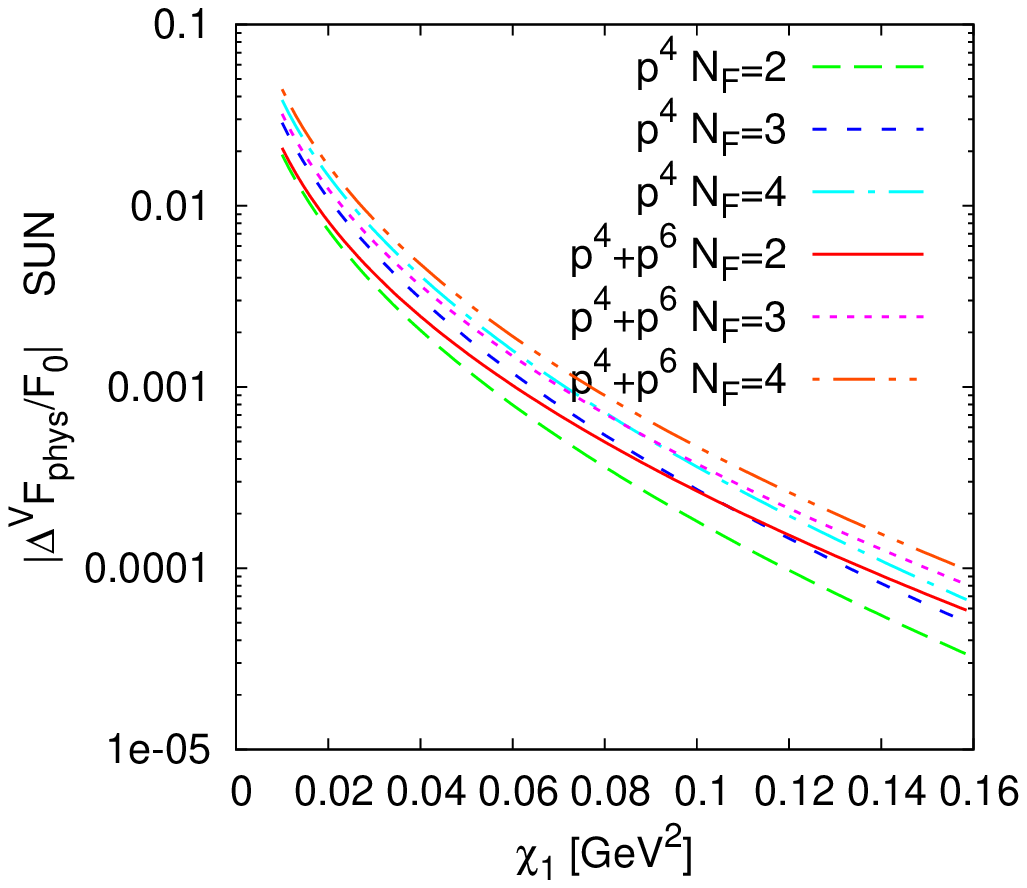}
\includegraphics[width=0.43\textwidth]{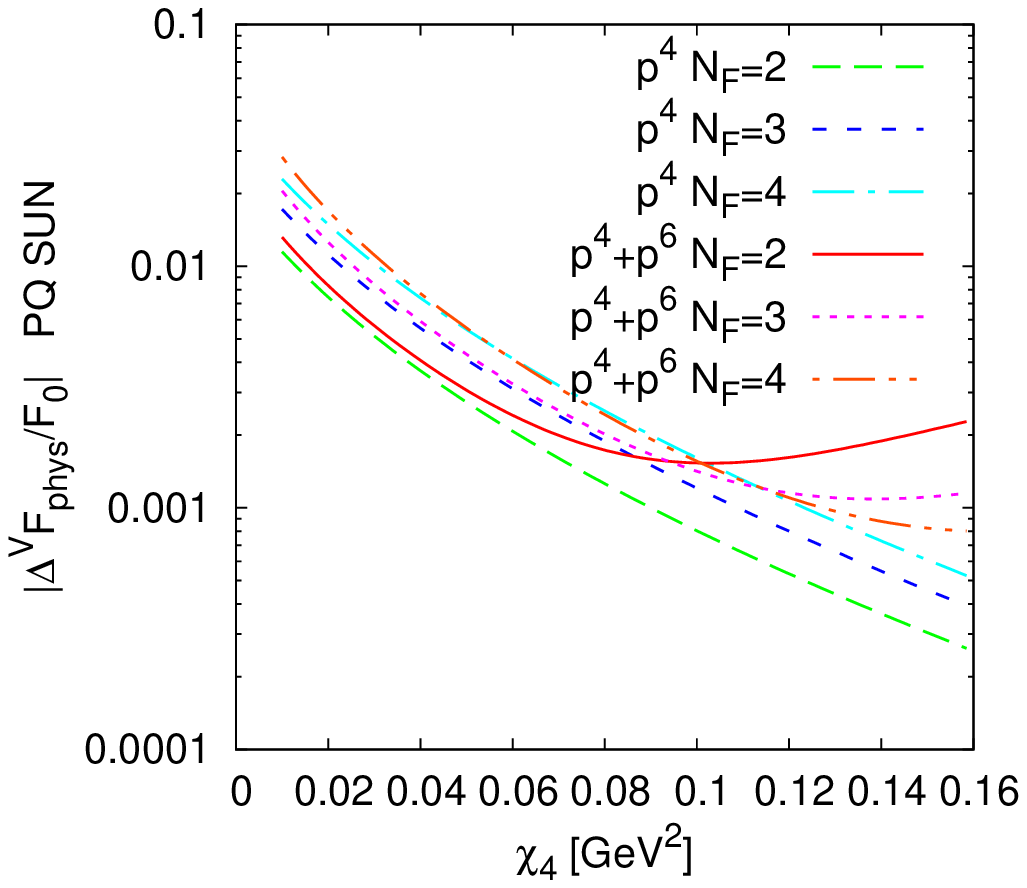}\\[-3mm]
\includegraphics[width=0.43\textwidth]{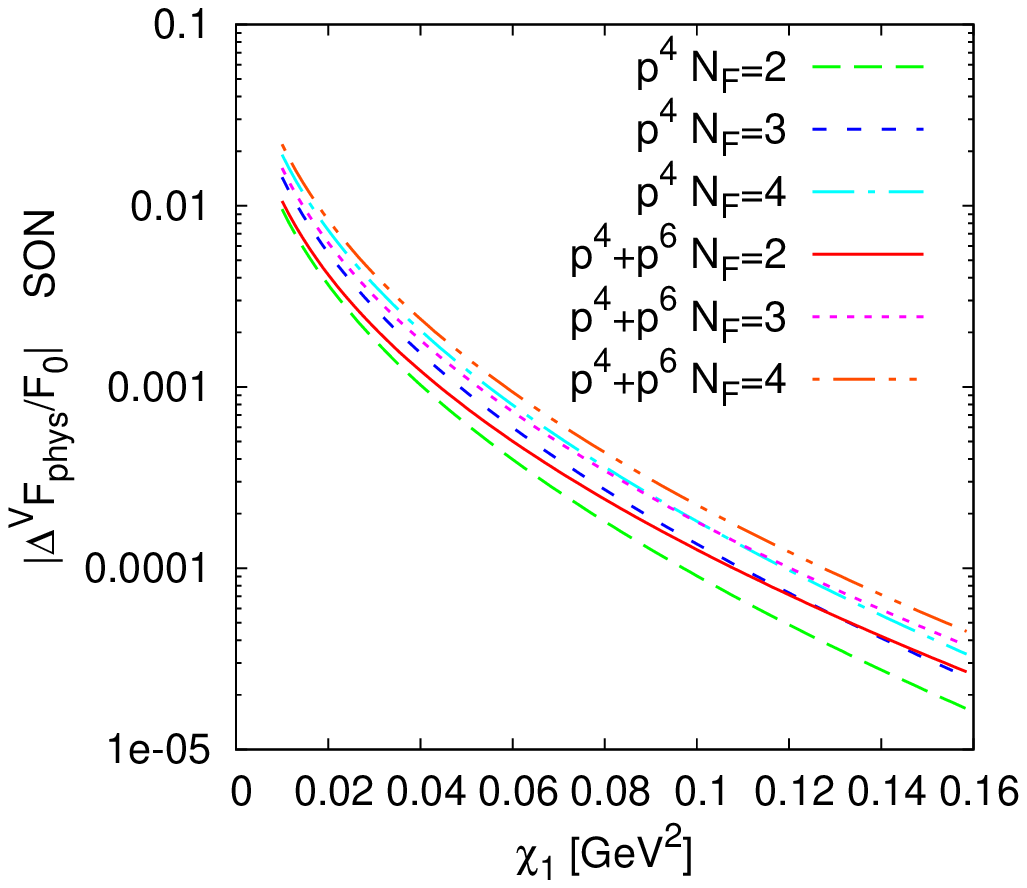}
\includegraphics[width=0.43\textwidth]{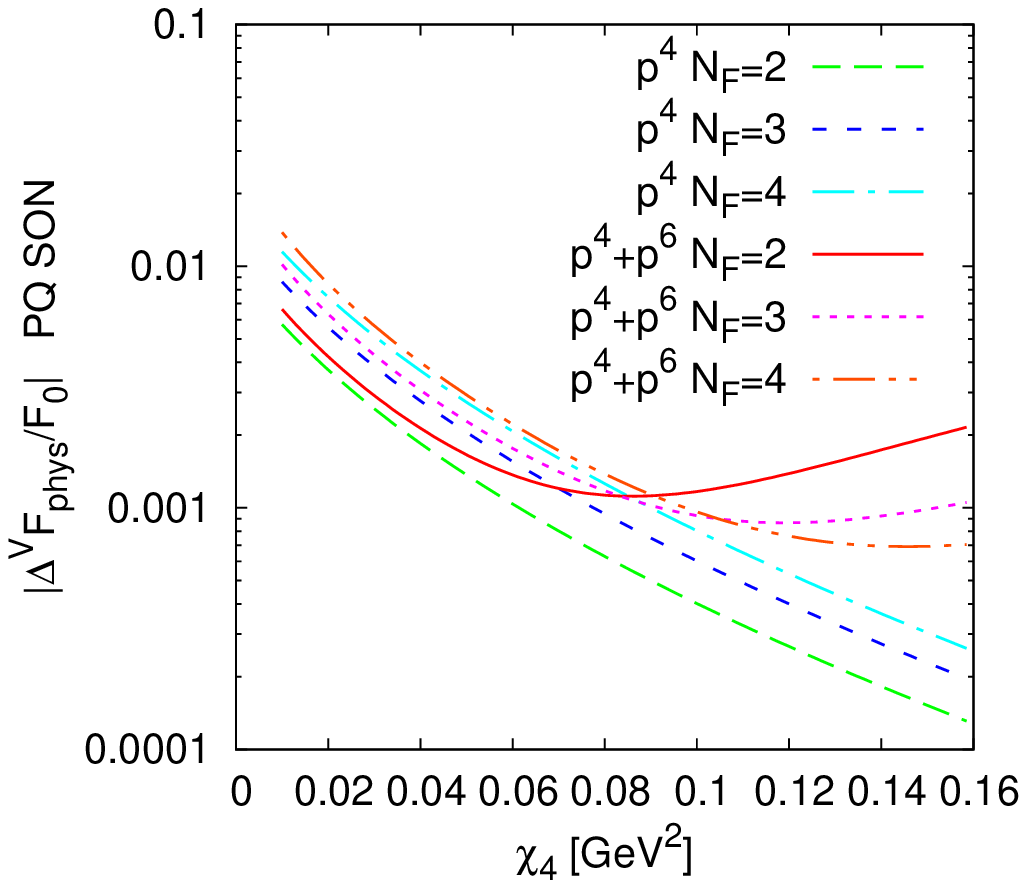}\\[-3mm]
\includegraphics[width=0.43\textwidth]{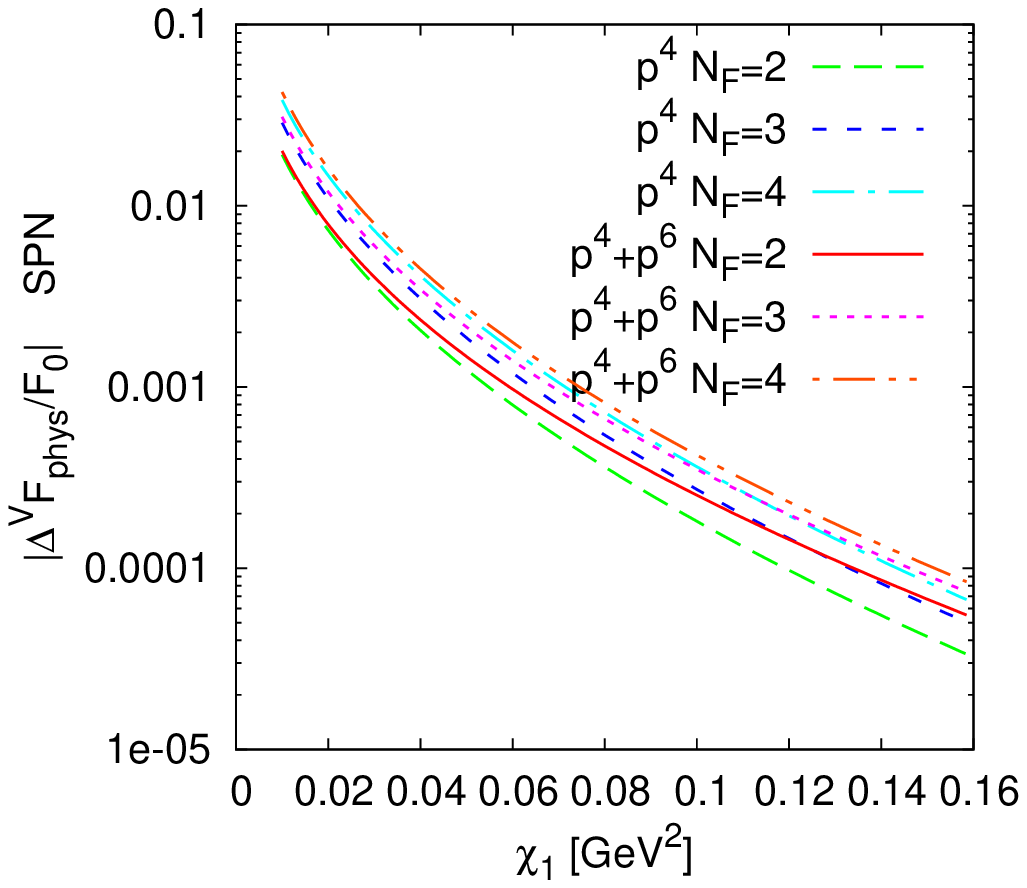}
\includegraphics[width=0.43\textwidth]{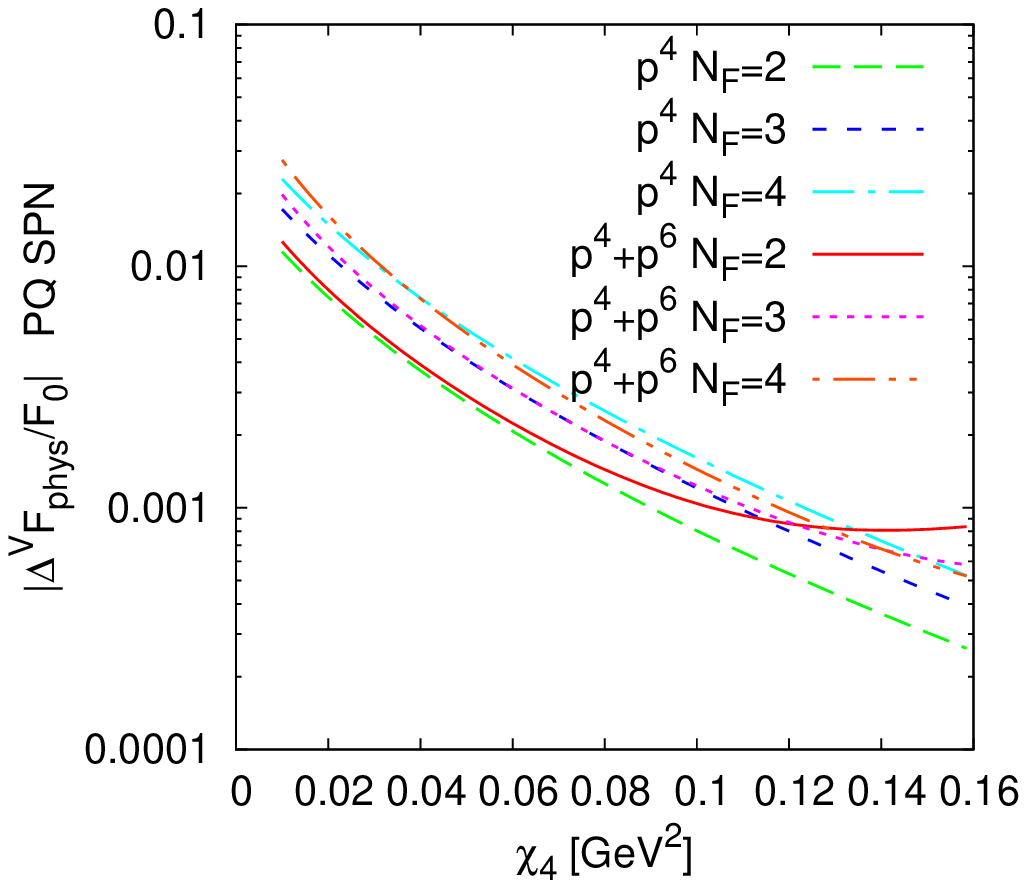}
\end{center}
\caption{\label{figdecayFV}
The absolute value of the finite volume
correction to the decay constant divided by the lowest order value $F_0=F_{LO}$
for the unquenched (left) as a function of $\chi_1$
and the partially quenched case (right) as a function of $\chi_4$
with $\chi_1=0.14^2$~GeV$^2$. 
Shown are the NLO ($p^4$) and NNLO ($p^4+p^6$)
results for three values of $N_F$.
Top line: $SU(N_F)\times SU(N_F)\to SU(N_F)$.
Middle line: $SU(N_F)\to SO(N_F)$.
Bottom line: $SU(2N_F)\to Sp(2N_F)$.
}
\end{figure}

\begin{figure}
\begin{center}
\includegraphics[width=0.43\textwidth]{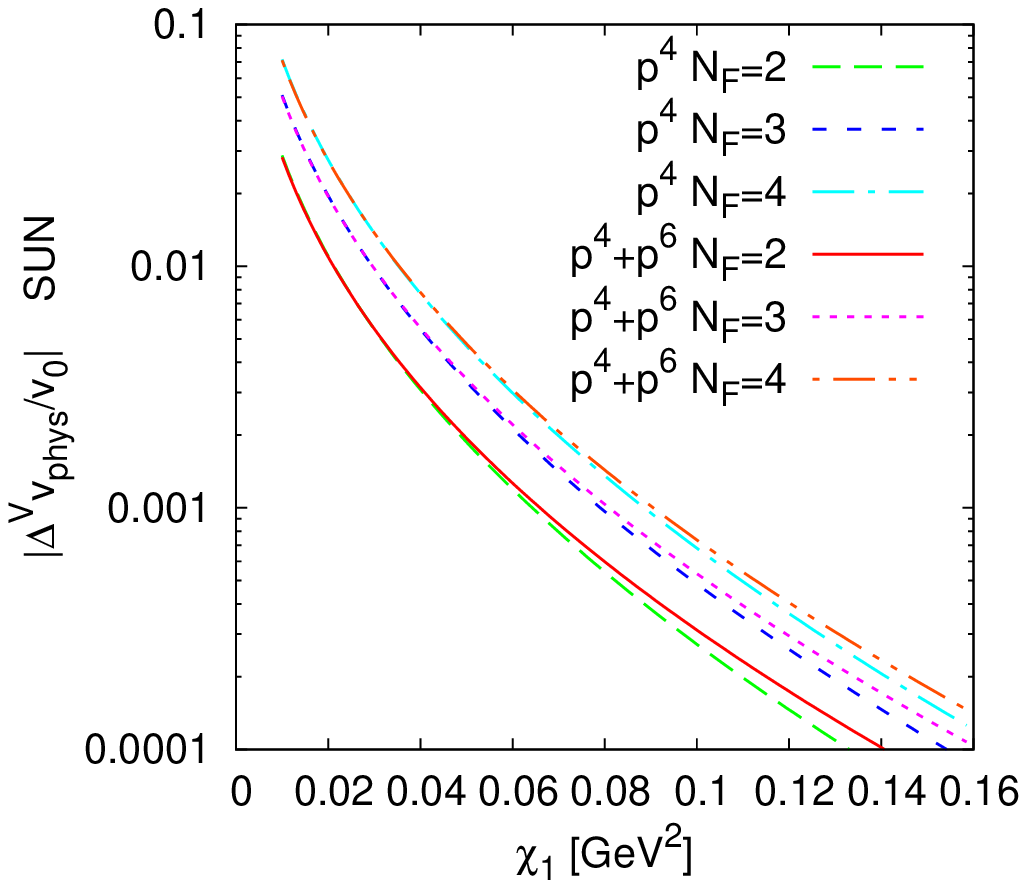}
\includegraphics[width=0.43\textwidth]{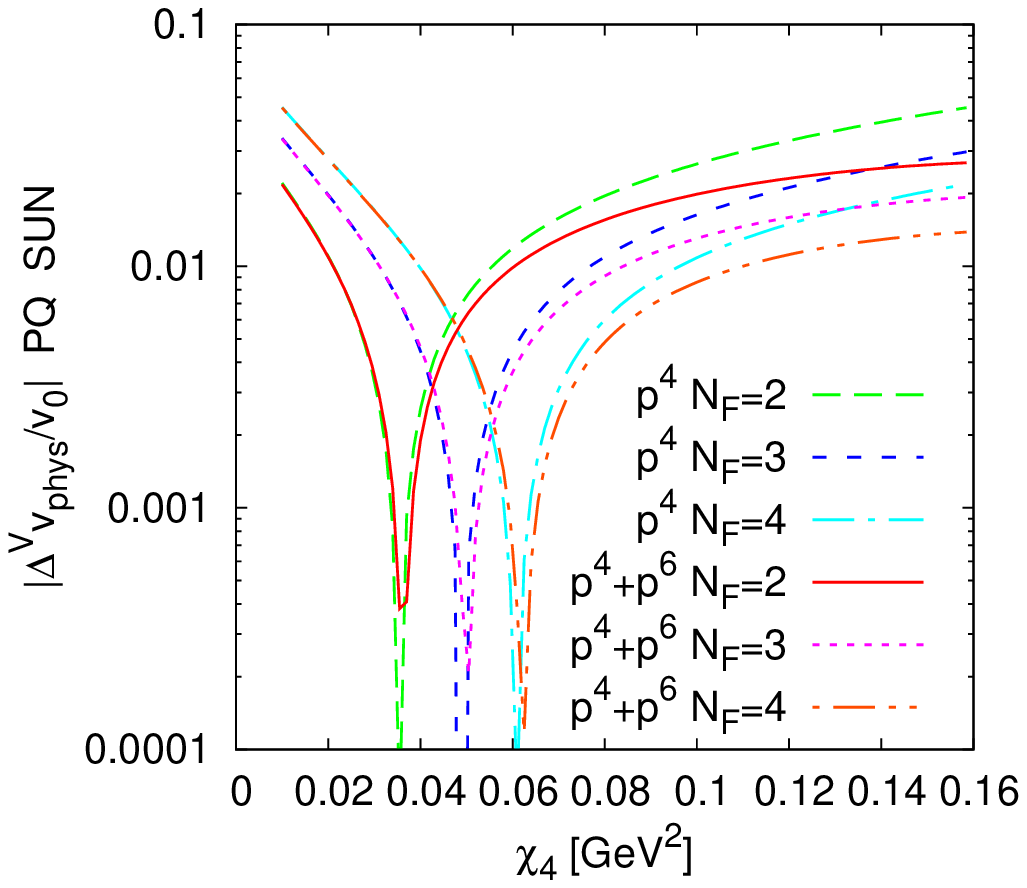}\\[-3mm]
\includegraphics[width=0.43\textwidth]{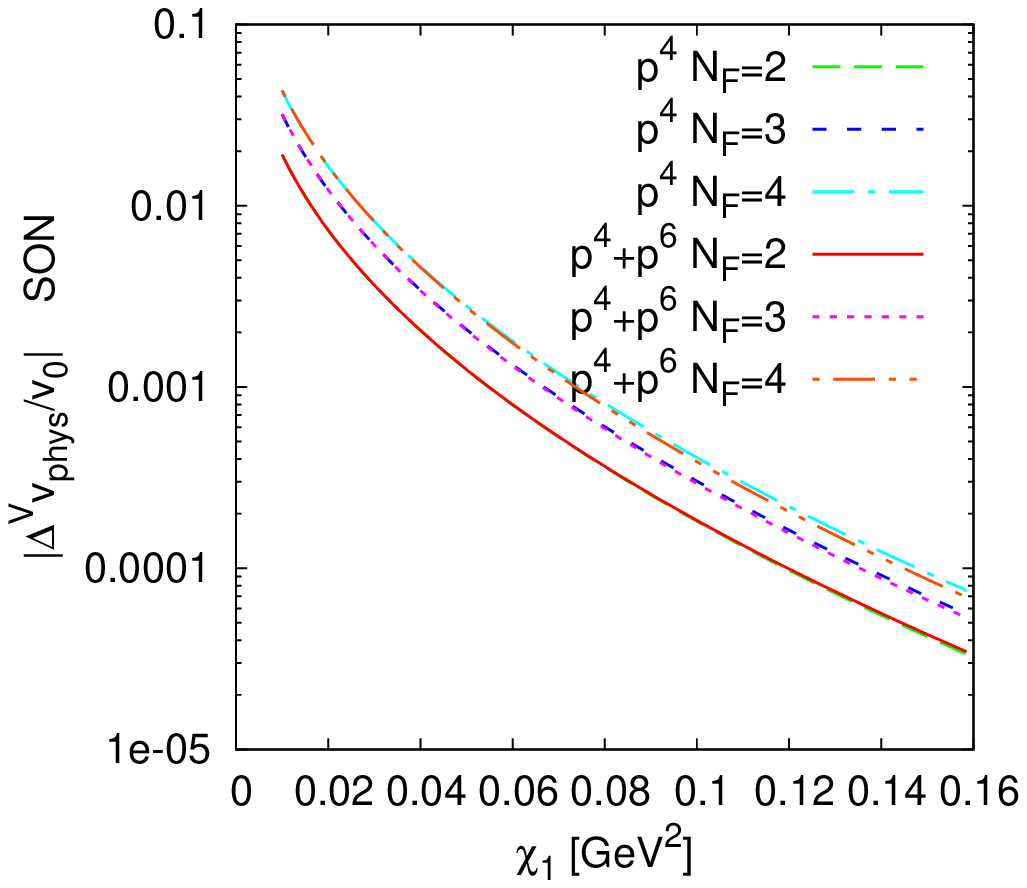}
\includegraphics[width=0.43\textwidth]{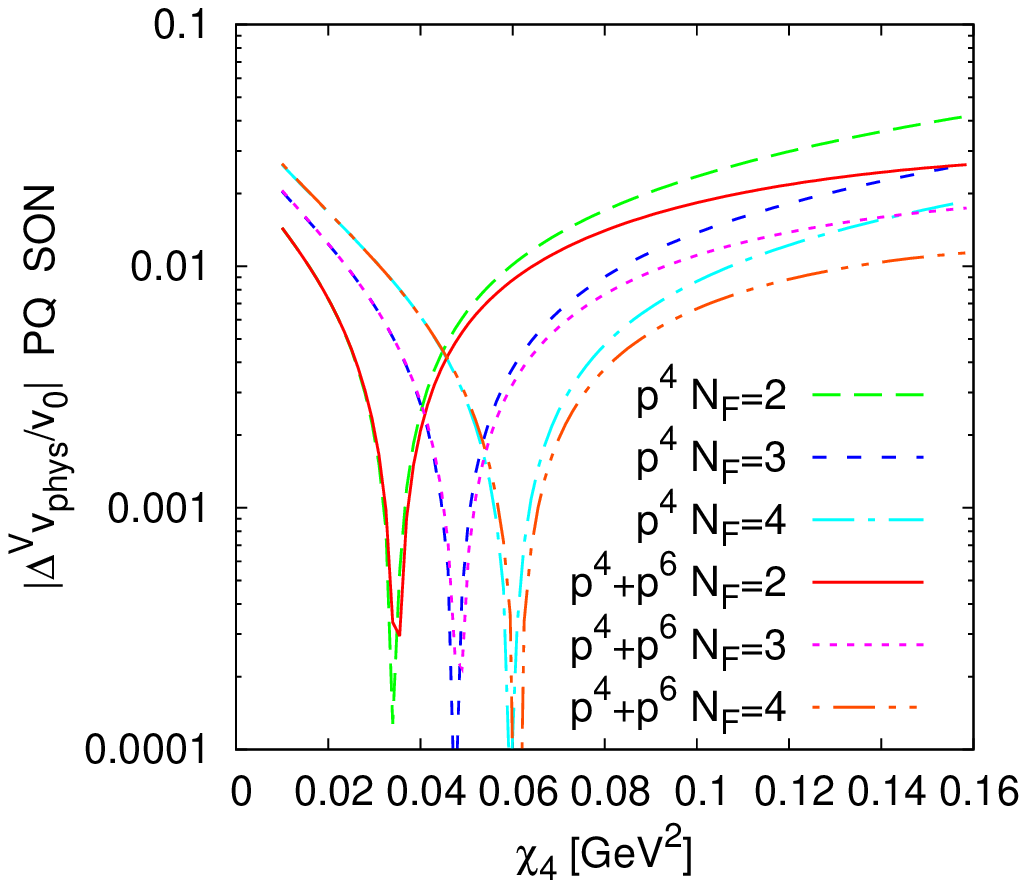}\\[-3mm]
\includegraphics[width=0.43\textwidth]{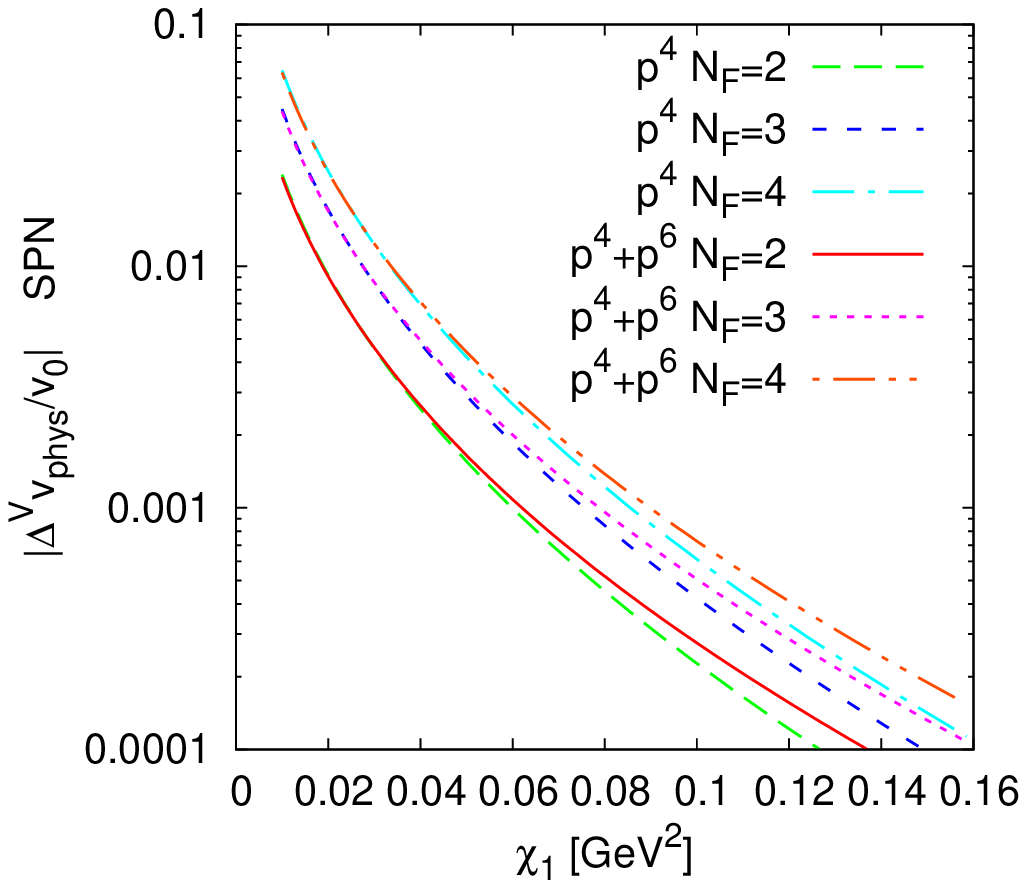}
\includegraphics[width=0.43\textwidth]{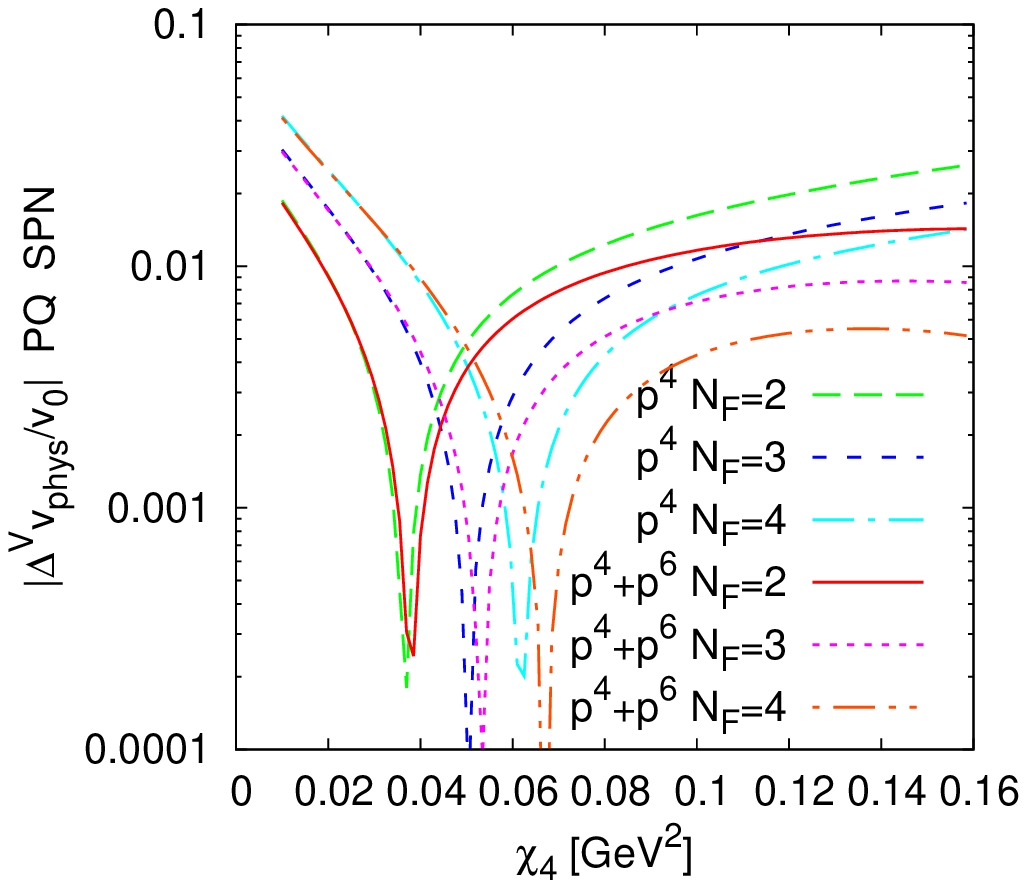}
\end{center}
\caption{\label{figvevFV}
The absolute value of the finite volume correction to the
vacuum expectation value divided by the lowest order value $v_0=v_{LO}$
for the unquenched (left) as a function of $\chi_1$
and the partially quenched case (right) as a function of $\chi_4$
with $\chi_1=0.14^2$~GeV$^2$. 
Shown are the NLO ($p^4$) and NNLO ($p^4+p^6$)
results for three values of $N_F$.
Top line: $SU(N_F)\times SU(N_F)\to SU(N_F)$.
Middle line: $SU(N_F)\to SO(N_F)$.
Bottom line: $SU(2N_F)\to Sp(2N_F)$.
}
\end{figure}

We can now make similar plots for the finite volume corrections.
The overall size of them is as expected.
The smallest $mL$ is about two for the left hand
sides of all plots. 
In the
unquenched case the exponential falloff with the mass is clearly visible.
The partially quenched cases contain a fixed mass scale $\chi_1$ which
is why the correction is more constant there, the stays
at the $mL=3$ point for the plots. The dips are caused by the
finite volume corrections going through zero.
The corrections to the mass are shown in Fig.~\ref{figmassFV},
the decay constant in Fig.~\ref{figdecayFV} and the vacuum expectation value
in Fig~\ref{figvevFV}.

\section{Conclusions}

We have calculated in the effective field theory for the three possible
symmetry breaking patterns the NNLO order finite volume and
partial quenching effects to NNLO in the expansion.
The results satisfy a large number of checks agreeing analytically and
numerically with earlier work that our results reduce to for some cases.
The analytical part of this work relied
heavily on \textsc{FORM} \cite{Vermaseren:2000nd}.

The analytical results are of reasonable length but given the total number
of results we have included them as FORM output in a supplementary file.
They can also be downloaded from \cite{chptweb}.

The numerical programs
have been included in \textsc{CHIRON} \cite{Bijnens:2014gsa} version 0.54
which can downloaded from \cite{chiron}. We have presented results in a number
of cases with typical QCD values of the parameters. The results are of the
expected sizes from earlier work in three flavour ChPT.
We hope these results will be useful for lattice studies of
these alternative symmetry breaking patterns.

\section*{Acknowledgements}

This work is supported in part by the Swedish Research Council grants
621-2011-5080 and 621-2013-4287. JB thanks the Centro de Ciencias
de Benasque Pedro Pascual, where part of this work was done, for hospitality.


\begin{thebibliography}{99}

\bibitem{Brambilla:2014jmp}
  N.~Brambilla {\it et al.},
  Eur.\ Phys.\ J.\ C {\bf 74} (2014) 10,  2981
  [arXiv:1404.3723 [hep-ph]].

\bibitem{talk1lattice2015}
F. Sannino, plenary talk at Lattice2015.

\bibitem{talk2lattice2015}
A. Hasenfratz, plenary talk at Lattice2015.

\bibitem{Kuti:2014epa}
  J.~Kuti,
  PoS LATTICE {\bf 2013} (2014) 004.

\bibitem{lattice}
  R.~Lewis, C.~Pica and F.~Sannino,
  Phys.\ Rev.\ D {\bf 85} (2012) 014504
  [arXiv:1109.3513 [hep-ph]].

  A.~Hietanen, R.~Lewis, C.~Pica and F.~Sannino,
  JHEP {\bf 1407} (2014) 116
  [arXiv:1404.2794 [hep-lat]].

  D.~Schaich {\it et al.} [LSD Collaboration],
  arXiv:1506.08791 [hep-lat].

  T.~N.~da Silva, E.~Pallante and L.~Robroek,
  arXiv:1506.06396 [hep-th].

  M.~P.~Lombardo, K.~Miura, T.~J.~Nunes da Silva and E.~Pallante,
  Int.\ J.\ Mod.\ Phys.\ A {\bf 29} (2014) 25,  1445007
  [arXiv:1410.2036 [hep-lat]].

  T.~Appelquist {\it et al.},
  Phys.\ Rev.\ D {\bf 85} (2012) 074505
  [arXiv:1201.3977 [hep-lat]].

  T.~DeGrand, Y.~Liu, E.~T.~Neil, Y.~Shamir and B.~Svetitsky,
  Phys.\ Rev.\ D {\bf 91} (2015) 114502
  [arXiv:1501.05665 [hep-lat]].

\bibitem{Andersen:2011yj}
  J.~R.~Andersen {\it et al.},
  Eur.\ Phys.\ J.\ Plus {\bf 126} (2011) 81
  [arXiv:1104.1255 [hep-ph]].

\bibitem{Techni1}
  F.~Sannino,
  Acta Phys.\ Polon.\  B {\bf 40} (2009) 3533
  [arXiv:0911.0931 [hep-ph]].

\bibitem{Techni2}
  C.~T.~Hill and E.~H.~Simmons,
  Phys.\ Rept.\  {\bf 381} (2003) 235
  [Erratum-ibid.\  {\bf 390} (2004) 553]
  [arXiv:hep-ph/0203079].

\bibitem{Weinberg:1978kz}
  S.~Weinberg,
  Physica A {\bf 96} (1979) 327.

\bibitem{Gasser:1983yg}
  J.~Gasser and H.~Leutwyler,
  Annals Phys.\  {\bf 158} (1984) 142.

\bibitem{Gasser:1984gg}
  J.~Gasser and H.~Leutwyler,
  Nucl.\ Phys.\ B {\bf 250} (1985) 465.

\bibitem{Peskin:1980gc}
  M.~E.~Peskin,
  Nucl.\ Phys.\  B {\bf 175} (1980) 197.

\bibitem{Preskill:1980mz}
  J.~Preskill,
  Nucl.\ Phys.\  B {\bf 177}, 21 (1981).

\bibitem{Dimopoulos:1979sp}
  S.~Dimopoulos,
  Nucl.\ Phys.\  B {\bf 168} (1980) 69.

\bibitem{Kogut}
  J.~B.~Kogut, M.~A.~Stephanov, D.~Toublan, J.~J.~M.~Verbaarschot and A.~Zhitnitsky,
  Nucl.\ Phys.\  B {\bf 582} (2000) 477
  [arXiv:hep-ph/0001171].

\bibitem{Splittorff}
  K.~Splittorff, D.~Toublan and J.~J.~M.~Verbaarschot,
  Nucl.\ Phys.\  B {\bf 620} (2002) 290
  [arXiv:hep-ph/0108040].

\bibitem{Toublan:1999hi}
  D.~Toublan and J.~J.~M.~Verbaarschot,
  Nucl.\ Phys.\ B {\bf 560} (1999) 259
  [hep-th/9904199].

\bibitem{Bijnens:2009qm}
  J.~Bijnens and J.~Lu,
  JHEP {\bf 0911} (2009) 116
  [arXiv:0910.5424 [hep-ph]].

\bibitem{Bijnens:2011fm}
  J.~Bijnens and J.~Lu,
  JHEP {\bf 1103} (2011) 028
  [arXiv:1102.0172 [hep-ph]].

\bibitem{Bijnens:2011xt}
  J.~Bijnens and J.~Lu,
  JHEP {\bf 1201} (2012) 081
  [arXiv:1111.1886 [hep-ph]].

\bibitem{Bernard:1993sv}
  C.~W.~Bernard and M.~F.~L.~Golterman,
  Phys.\ Rev.\ D {\bf 49} (1994) 486
  [hep-lat/9306005].

\bibitem{Bernard:2013kwa}
  C.~Bernard and M.~Golterman,
  Phys.\ Rev.\ D {\bf 88} (2013) 1,  014004
  [arXiv:1304.1948 [hep-lat]].

\bibitem{Sharpe:2000bc}
  S.~R.~Sharpe and N.~Shoresh,
  Phys.\ Rev.\ D {\bf 62} (2000) 094503
  [hep-lat/0006017].

\bibitem{Sharpe:1992ft}
  S.~R.~Sharpe,
  Phys.\ Rev.\ D {\bf 46} (1992) 3146
  [hep-lat/9205020].

\bibitem{Bijnens:2004hk}
  J.~Bijnens, N.~Danielsson and T.~A.~Lahde,
  Phys.\ Rev.\ D {\bf 70} (2004) 111503
  [hep-lat/0406017].

\bibitem{Bijnens:2005ae}
  J.~Bijnens and T.~A.~Lahde,
  Phys.\ Rev.\ D {\bf 71} (2005) 094502
  [hep-lat/0501014].

\bibitem{Bijnens:2006jv}
  J.~Bijnens, N.~Danielsson and T.~A.~Lahde,
  Phys.\ Rev.\ D {\bf 73} (2006) 074509
  [hep-lat/0602003].

\bibitem{Gasser:1986vb}
  J.~Gasser and H.~Leutwyler,
  Phys.\ Lett.\ B {\bf 184} (1987) 83.

\bibitem{Gasser:1987ah}
  J.~Gasser and H.~Leutwyler,
  Phys.\ Lett.\ B {\bf 188} (1987) 477.

\bibitem{Gasser:1987zq}
  J.~Gasser and H.~Leutwyler,
  Nucl.\ Phys.\ B {\bf 307} (1988) 763.

\bibitem{Colangelo:2006mp}
  G.~Colangelo and C.~Haefeli,
  Nucl.\ Phys.\ B {\bf 744} (2006) 14
  [hep-lat/0602017].

\bibitem{Bijnens:2006ve}
  J.~Bijnens and K.~Ghorbani,
  Phys.\ Lett.\ B {\bf 636} (2006) 51
  [hep-lat/0602019].

\bibitem{Damgaard:2008zs}
  P.~H.~Damgaard and H.~Fukaya,
  JHEP {\bf 0901} (2009) 052
  [arXiv:0812.2797 [hep-lat]].

\bibitem{Bijnens:2013doa}
  J.~Bijnens, E.~Boström and T.~A.~Lähde,
  JHEP {\bf 1401} (2014) 019
  [arXiv:1311.3531 [hep-lat]].

\bibitem{Bijnens:2014dea}
  J.~Bijnens and T.~Rössler,
  JHEP {\bf 1501} (2015) 034
  [arXiv:1411.6384 [hep-lat]].

\bibitem{Bijnens:2015dra}
  J.~Bijnens and T.~Rössler,
  arXiv:1508.07238 [hep-lat].

\bibitem{analyticalresults} See the supplementary file analyticalresults.txt.
This can also be downloaaded from \cite{chptweb}.

\bibitem{Bijnens:2014gsa}
  J.~Bijnens,
  Eur.\ Phys.\ J.\ C {\bf 75} (2015) 1,  27
  [arXiv:1412.0887 [hep-ph]].

\bibitem{chiron}  \href{http://www.thep.lu.se/\%7Ebijnens/chiron/}
                       {http://www.thep.lu.se/\textasciitilde{}bijnens/chiron/}

\bibitem{Bijnens:1999sh}
  J.~Bijnens, G.~Colangelo and G.~Ecker,
  JHEP {\bf 9902} (1999) 020
  [hep-ph/9902437].

\bibitem{Bijnens:1999hw}
  J.~Bijnens, G.~Colangelo and G.~Ecker,
  Annals Phys.\  {\bf 280} (2000) 100
  [hep-ph/9907333].

\bibitem{CCWZ}
  S.~R.~Coleman, J.~Wess and B.~Zumino,
  Phys.\ Rev.\  {\bf 177} (1969) 2239;
  C.~G.~.~Callan, S.~R.~Coleman, J.~Wess and B.~Zumino,
  Phys.\ Rev.\  {\bf 177} (1969) 2247.

\bibitem{Sharpe:2001fh}
  S.~R.~Sharpe and N.~Shoresh,
  Phys.\ Rev.\ D {\bf 64} (2001) 114510
  [hep-lat/0108003].

\bibitem{Damgaard:2000gh}
  P.~H.~Damgaard and K.~Splittorff,
  Phys.\ Rev.\ D {\bf 62} (2000) 054509
  [hep-lat/0003017].

\bibitem{Levinsen:2003be}
  J.~Levinsen,
  Phys.\ Rev.\ D {\bf 67} (2003) 125009
  [hep-th/0301008].

\bibitem{Bijnens:2014lea}
  J.~Bijnens and G.~Ecker,
  Ann.\ Rev.\ Nucl.\ Part.\ Sci.\  {\bf 64} (2014) 149
  [arXiv:1405.6488 [hep-ph]].

\bibitem{Amoros:1999dp}
  G.~Amor\'os, J.~Bijnens and P.~Talavera,
  Nucl.\ Phys.\ B {\bf 568} (2000) 319
  [hep-ph/9907264].


\bibitem{Amoros:2000mc}
G.~Amor\'os, J.~Bijnens and P.~Talavera,
{\em Nucl. Phys.} B { 585} (2000) 293
[Erratum-ibid.\ B { 598} (2001) 665]
[hep-ph/0003258].

\bibitem{Vermaseren:2000nd}
  J.~A.~M.~Vermaseren,
  arXiv:math-ph/0010025.

\bibitem{chptweb}  \href{http://www.thep.lu.se/\%7Ebijnens/chpt/}
                       {http://www.thep.lu.se/\textasciitilde{}bijnens/chpt/}


\end{thebibliography}
\end{document}